\documentclass[twocolumn]{aastex631}

\begin{document}

\title{Extragalactic Stellar Streams in Time-Dependent Cosmological Halos}

\author[0000-0003-0256-5446]{Sarah Pearson}
\affiliation{DTU Space, Technical University of Denmark, Elektrovej 327, DK- 2800 Kgs. Lyngby, Denmark}
\affiliation{DARK, Niels Bohr Institute, University of Copenhagen, Jagtvej 155A, 2200 Copenhagen,  Denmark}
\correspondingauthor{Sarah Pearson}
\email{sapea@dtu.dk}

\author[0000-0001-8042-5794]{Jacob Nibauer}
\affiliation{Center for Astrophysics $\vert$ Harvard \& Smithsonian, 60 Garden Street, Cambridge, MA 02138, USA}
\affiliation{Department of Astrophysical Sciences, Princeton University, Princeton, NJ 08544, USA}

\author[0000-0002-6993-0826]{Emily C. Cunningham}
\affiliation{Department of Astronomy, Boston University, 725 Commonwealth Avenue, Boston, MA 02215, USA}

\author[0000-0003-0872-7098]{Adrian~M.~Price-Whelan}
\affiliation{Center for Computational Astrophysics, Flatiron Institute, 162 Fifth Ave, New York, NY 10010, USA}

\author[0000-0001-7928-1973]{Adrien C. R. Thob}
\affiliation{Department of Physics and Astronomy, University of Pennsylvania, 209 South 33rd Street, Philadelphia, PA 19104 USA}

\author[0000-0002-8354-7356]{Arpit Arora}
\affiliation{Department of Astronomy and DiRAC Institute, University of Washington, 3910 15th Ave NE, Seattle, WA, 98195, USA}

\author[0000-0003-3939-3297]{Robyn E. Sanderson}
\affiliation{Department of Physics \& Astronomy, University of Pennsylvania, 209 S 33rd Street, Philadelphia, PA 19104, USA}

\begin{abstract}
Upcoming and ongoing surveys will detect thousands of stellar streams around  galaxies other than the Milky Way. Studies from the Milky Way 
have shown that time-dependent evolution of the Galactic halo plays a key role in shaping stellar streams, but remains unexplored for extragalactic stellar streams. 
We use the FIRE-2 m12m cosmological zoom-in simulation, mock observed as an extragalactic system including three stellar streams, to examine how halo time-dependence affects progenitor and host halo inference from extragalactic systems. 
We show that two of the three m12m streams are well reproduced in a static halo if we only allow for tidal stripping near pericenter. 
We apply the extragalactic stream fitting code \texttt{X-Stream} to each  mock observed stream, and obtain constraints on the host dark matter halo and stream properties. Using on-sky morphology alone and then fixing the progenitor radial velocities, we compare recovered orbits and halo parameters to the FIRE-2 m12m ground truth.
For the longest stream with a looped segment, we find unbiased strong constraints on progenitor and halo properties. 
For the shortest stream, we find limits on orbital parameters, but no constraints on progenitor and halo mass unless we include fainter, more extended debris. 
For the most massive stream, which was not well produced in a static halo, the recovered orbital parameters are biased,  reflecting unmodeled time-dependence. 
 We conclude that imaging of stream debris from extragalactic dwarf galaxies can, in some cases, be used to infer present-day dark matter halo properties, even in a cosmological environment.
\end{abstract}

\keywords{galaxies: dark matter halos, galaxies: kinematics and dynamics, galaxies: stellar streams}

\section{Introduction} \label{sec:intro}
Stellar streams form when stars tidally strip from a progenitor system orbiting another galaxy \citep{Johnston1996}. Streams are sensitive to  time-dependent processes, such as perturbations from the Galactic bar, dark matter subhalos, giant molecular clouds, spiral arms, and the tilting of the disk \citep{erkal2017,pearson2017,yoon2011,price-whelan2018,amorisco2016,banik2019,nibauer2024}. 
Individual orbits of stream progenitors can change over time due to time-dependent effects in the halo (e.g., \citealt{arora2022,Santistevan2024, nibauer2024}), and stream morphologies can be affected by large-scale features of the host and by close encounters with other accreted satellites
\citep{dillamore2022, woudenberg2023, Foote2025, arora2026}. Halo mass growth can affect stream morphologies  \citep{buist2015}, and the accretion of major satellites, such as the Large Magellanic Cloud (LMC), can affect both progenitor orbits and individual stream star trajectories \citep{erkal2019,vasiliev2021,shipp2021,lilleengen2023,brooks2025,sachi2025}. 

More than 100 candidate streams have been found throughout the Milky Way \citep[e.g.,][]{bonaca2025}, most of which likely formed from disrupted star clusters.
Stellar streams in extragalactic systems beyond the Milky Way will generally be detected from more massive progenitors and further out in the halos (Pérez-Herrero, in prep.). Such streams will be less sensitive to the inner halo perturbations, however, they can be affected by 
cosmological growth from mergers and accretion, which shape the evolution of galaxies throughout the history of the Universe. 
Targeted observations and broader surveys of low surface brightness features have discovered hundreds of stellar streams from tidally disrupted dwarf galaxies  around galaxies other than the Milky Way \citep{malin1997,ibata2001,denja2016,delgado2023,miro2023, miro2024b, fielder2025,sola2025,Koblischke2025}, and ongoing and upcoming surveys will find many more (e.g. {\it Euclid}: \citealt{racca2016,starkman2026}, {\it ARRAKIHS}: \citealt{guzman2022}, the Vera Rubin Observatory: \citealt{ivezic2019}, the {\it Nancy Grace Roman Space Telescope}: \citealt{spergel2015}, and LIGHTS:  \citealt{Zaritsky2024,Zaritsky2026A}). 
These extragalactic stellar streams have already been used to  map accretion histories of other galaxies \citep[e.g.,][]{denja2016} and to test differences between observed and theoretical predictions for stellar stream frequencies  \citep{miro2024a}. 

In the Milky Way, stellar streams have been used to map dark matter halo properties such as its mass distribution, flattening, and radial profile \citep[e.g.,][]{Law2010,koposov2010,veraciro2013,belokurov2014, kuepper2015,bovy2016,bonaca2018}.
Recent theoretical work has demonstrated that we can also use extragalactic stellar streams to learn about the properties of their hosts' dark matter halos, the stream progenitors' orbits, and masses \citep{Fardal2013,foster2014,pearson2022b,nibauer2023,walder2025,NibPear2025,Chemaly2026,Wu2026,starkman2026,Chemaly2026b}. 
\citet{dokkum2019} showed that the morphologies of extragalactic streams can be reproduced with stream models evolved in static galactic potentials \citep[see also][]{foster2014,amorisco2015}. 
With access to kinematic and morphological data of the Giant Southern Stream in M31, \citet{Fardal2013} used a Bayesian analysis to constrain the properties of both the host and the stream. More recently, \citet{pearson2022b} developed a stream fitting technique with particle-spray simulations \citep{Fardal2015}, for 10 different NFW halos \citep{nfw1997}. They could constrain the host halo mass of Centaurus A by modeling its straight stream if they fixed the radial velocity and stellar mass of its progenitor from observations.  

The morphologies of extragalactic streams alone can also inform halo properties such as halo flattening and density slopes \citep{nibauer2023,walder2025, NibPear2025,Chemaly2026,Wu2026}, which can be used as direct cosmological tests at the stream population level \citep[][]{starkman2026,Chemaly2026b}. 

 \begin{deluxetable*}{lcccccccc}[t]
\tablecaption{True present-day FIRE stream 6D parameters}
\tablecolumns{9}
\tablenum{1}\label{tab:streams}
\tablewidth{0pt}
\tablehead{\colhead{\bf Stream } & 
\colhead{\bf $x$ }  &{ $y$ (los) }    &{\bf $z$ }  &{\bf $v_x$ }  &{ $v_y$ (los) }  &{\bf $v_z$ }  &{\bf $m_{prog}$$^a$ } &{\bf $t_{\rm accreted}$$^b$}\\
\colhead{\bf } & 
\colhead{ [kpc] }  &{ [kpc] }    &{ [kpc] }  &{[km/s] }  &{ [km/s] }  &{ [km/s] }  &{ 10$^{8}$ M$_{\odot}$ } &{ Gyr}
}
\startdata
{\bf Massive}  & 62.1 &   26.0& 66.4& -68.2&  124 & 50.8 & 11.0 &  $-7.17$ \\
{\bf S-shaped}  &$-68.3$ &90.6 &24.6&  $-129$& 74.5 &$-91.8$ & 3.4 & $-7.17$ \\
{\bf Curlicue}  & $-11.9$ &  $12.6$ & $72.9$ & $-90.3$ &$116$  & $-10.7$& 6.1 & $-8.25$   \\
\enddata
\tablenotetext{}{$^a$ The stellar mass is listed as the total stellar mass of the present-day progenitor (if intact) and stellar mass in its stream. $^b$ Time of first and last crossing of the virial radius since present day ($t=0$), which is the upper limit for when the stream could start forming. } 
\end{deluxetable*}

Despite growing interest in halo time-dependence for Milky Way streams, extragalactic stream fitting has largely ignored this effect. 
Extragalactic streams could constrain the statistical properties of dark matter halos at various redshifts, such as their shapes, masses, and radial profiles, testable against theoretical predictions and different dark matter models \citep[e.g.,][]{Allgood2006,Chua2019,Despali2025}. However, understanding how the cosmological build-up and evolution of a halo can shape the observable properties of external stream remains underexplored. %

In this paper, we mock observe the FIRE-2 m12m cosmological zoom-in simulation at $z=0$ from the \textit{Latte} suite and treat it as an extragalactic system. We determine whether a static potential with a disk, bulge, and spherical halo is capable of reproducing the mock observations, and explore biases in inference of the host halo properties from  stream morphology. 
For inference, we apply the \citet{NibPear2025} \texttt{X-Stream} code, which uses generative models to translate stellar stream imaging into constraints on stream progenitors and host dark matter halos.  We compare our inference to the true stream progenitor properties and to the halo properties of m12m, using the present-day stream morphologies alone and then folding in radial velocity information.

The paper is summarized as follows. In Section \ref{sec:methods}, we describe the FIRE-2 m12m simulation, its streams and describe our application of the  \texttt{X-Stream} sampler. In Section \ref{sec:Results}, 
we show results on stream morphology, orbital constraints, as well as constraints on the radial dark matter profile and progenitor and halo mass.  In Section \ref{sec:discuss}, we discuss our results in the context of future stream observations, and we conclude our findings in Section \ref{sec:conclusion}.

\section{Methods} \label{sec:methods}
In this Section, we first introduce the FIRE-2 m12m simulation and its streams (Section \ref{sec:FIRE}), and then describe how the simulation is mock-observed (Section \ref{sec:euclid}). In Section \ref{sec:potential}, we extract the ground truth m12m potential properties from fits to the present-day FIRE particle data, and we summarize our application of the \texttt{X-Stream} sampler in Section \ref{sec:xstream}. 

\subsection{FIRE-2 m12m} \label{sec:FIRE}
In this work, we make use of the m12m simulation from the \textit{Latte} suite of FIRE-2 cosmological zoom-in simulations of MW-mass galaxies (first introduced in \citealt{Wetzel2016}). These simulations are publicly available \citep{wetzel2023public, wetzel2025second}\footnote{At \url{http://flathub.flatironinstitute.org/fire}}.
The m12m simulation was first introduced in \citealt{Hopkins2018}. 
This simulation was run using the GIZMO\footnote{\url{http://www.tapir.caltech.edu/~phopkins/Site/GIZMO.html}} gravity plus hydrodynamics code in meshless finite-mass (MFM) mode (\citealt{Hopkins2015}) with the FIRE-2 physics model (\citealt{Hopkins2018}).  
We refer the reader to the above papers for more details about the FIRE-2 implementation. 

The FIRE-2 halos selected for the \textit{Latte} suite of simulations are isolated galaxies at present day, with masses in the range of $M_{200}=1-2 \times 10^{12}~M_{\odot}$, similar to the MW and M31. 
The star particles' initial masses are $\sim 7000 M_{\odot}$, while the average $z=0$ particle mass is $\sim 5000~M_{\odot}$, as a result of stellar mass loss (the specific star particle mass depends on age). In this study, we focus on  m12m with $M_{\rm 200} \sim 1.58\times 10^{12}$ M$_{\odot}$, $M_{\rm *, 90} \sim 1.12\times 10^{11}$ M$_{\odot}$. m12m hosts the oldest and the largest disk in the \textit{Latte} suite \citep{Sanderson2020}.
m12m is also one of the most isolated simulations with no major mergers ($>1:50$) and the fewest changes to its profile and shape during the last 5 Gyr \citep{Panithanpaisal2021,Horta2023, arora2025}.

The stellar streams studied in this work were first identified in \citet{Panithanpaisal2021} and \citet{Horta2023}, and their properties are also discussed in \citet{Cunningham2022} and \citet{Shipp2023}. To identify the properties of the dwarf galaxy progenitors of these systems, we use the halo catalogs for each snapshot (created using the ROCKSTAR 6D halo finder;  \citealt{Behroozi2013}), as well as the merger trees, which connect the halo catalogs across time (created by CONSISTENT-TREES; \citealt{Behroozi2013b}). To identify all star particles associated with a given disrupted dwarf galaxy, we use the merger trees to track the star particles associated with the subhalo over time. Each snapshot is separated in time by $\sim25$ Myr. Using every fifth snapshot, we take all star particles that are assigned to the subhalo, until the subhalo is disrupted and no longer tracked by ROCKSTAR. This selection method can result in contamination from disk stars (depending on how close to the disk the dwarf galaxy passes to the disk during its evolution), which can be manually removed in post-processing based on their ages and formation distances (see \citealt{Horta2023} for details). 

\begin{figure*}
    \centering
    {\includegraphics[width=0.95\textwidth]{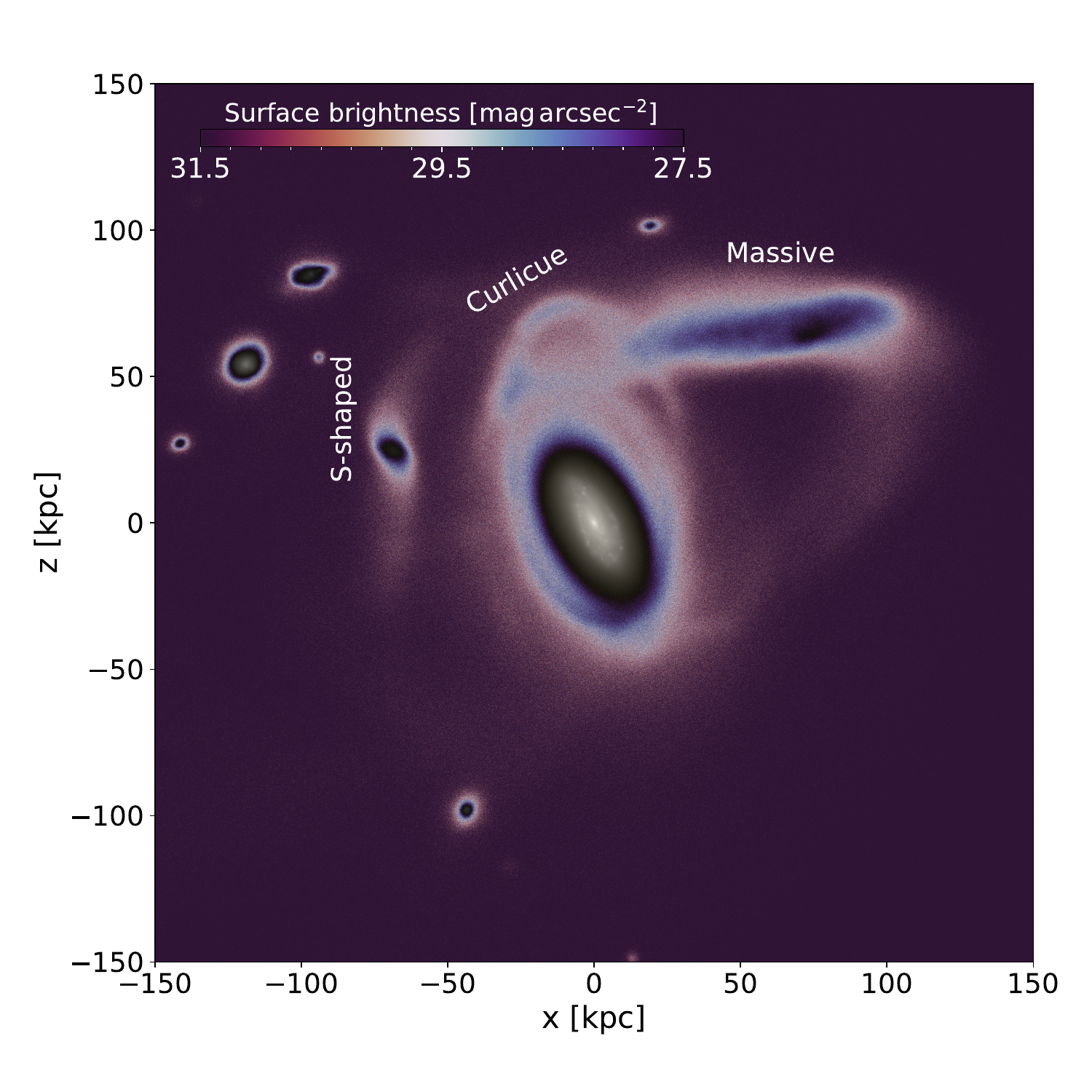}}
    \caption{
    Three stellar streams of m12m are visible in the Euclid Wide and Deep surveys when placed at Centaurus A's distance ($\sim3.8$ Mpc,  \citealt{Harris2010}). This mock observation shows integrated surface brightness images of m12m's stellar content, generated with the \texttt{py-ananke} synthetic survey pipeline \citep{Thob2024}. The brighter regime shows an RGB composite, with red, green and blue channels encoding Euclid's NISP/J, NISP/Y and VIS fluxes, respectively. The fainter regime shows the VIS image alone, with color scaling matched to the surface brightness limits of Euclid's Wide and Deep surveys (29.5 and 31.5 $\rm{mag} \, \rm{arcsec}^{-2}$, respectively; \citealt{EuclidXVI2022}).}
    \label{fig:sim}
\end{figure*}

For each of the three streams, we locate their progenitor position at present day. Since the two of the streams have fully disrupted progenitors, we define the progenitor position from the median positions and velocities for the 500 particles closest to highest density location in the unwrapped streams. This gives us access to the streams' present-day 6D phase space coordinates, which we list in Table \ref{tab:streams}.

\subsection{Mock observing m12m} \label{sec:euclid}
In Figure \ref{fig:sim}, we show a mock observation of the m12m simulation $z=0$ snapshot and highlight three stellar streams, which we label the ``Massive stream'', the ``S-shaped stream'', and the ``Curlicue stream''. The Curlicue stream crosses the virial radius once, 8.25 Gyr before present day. The other two streams cross it once,  7.17 Gyr before present day. 
We produced the surface brightness map by integrating the photometry of a synthetic star survey generated with \texttt{py-ananke} \citep{Thob2024} using the m12m star particles as input. \texttt{py-ananke} uses isochrone interpolation with phase-space density and IMF sampling to split the input particles from first principles into a consistent catalog of individual mock stars, assigning their corresponding astrometry and {\it Euclid} photometry. 
Figure \ref{fig:sim} displays this catalog by showing in the brighter regions (VIS surface brightness $> 27.5 ~\rm{mag} \, \rm{arcsec}^{-2}$) a RGB composite of maps where the red, green and blue channels encode {\it Euclid}’s NISP/J, NISP/Y and VIS fluxes, respectively. In the deep fainter regime, the color scaling reflects the surface brightness in {\it Euclid}'s VIS filter, with a colormap profile chosen to match the limits determined by \citet{EuclidXVI2022} for {\it Euclid}'s Wide and Deep surveys (29.5 and 31.5 $\rm{mag} \, \rm{arcsec}^{-2}$, respectively). 
To treat m12m as an extragalactic system, we assume that we know the distance to the host galaxy and transform into the galactocentric rest frame of m12m. For a discussion of the assumption of known host distance see \citet{NibPear2025}. 
Here  $x,z$ denotes the projected sky plane,  m12m is centered at ($x,z$) = (0,0), and $y$ denotes the line of sight direction. 

\begin{figure}[b]
    \centering
    \includegraphics[width=\columnwidth]{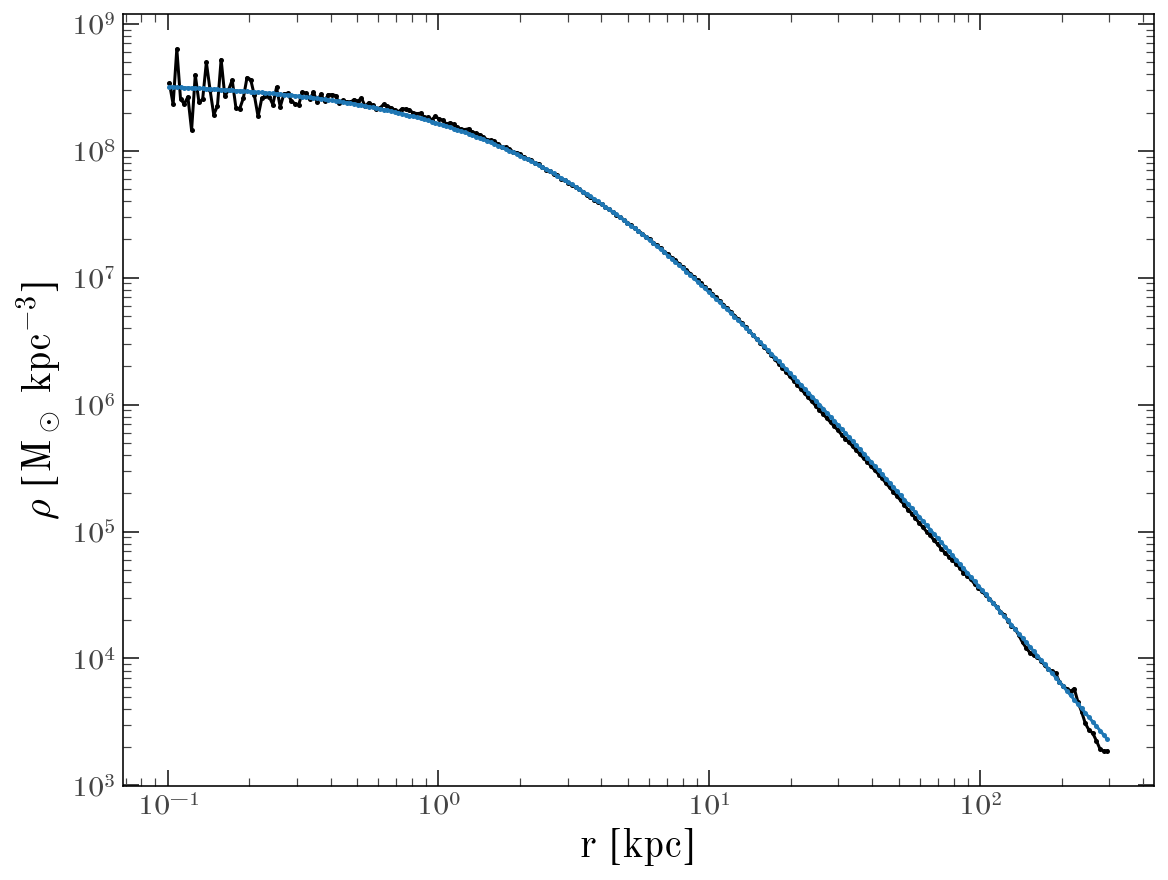}
    \caption{Total density of dark matter as a function of galactocentric radius from the FIRE m12m particle data at $z=0$ (black) compared to the density profile from our analytic Zhao profile fit  (blue). The cored center of m12m in dark matter was also identified in \citet{Vargya2022, ou2025}. } 
    \label{fig:zhao_z0}
\end{figure}

\subsection{Fitting the present-day potential of m12m} \label{sec:potential}
When we observe extragalactic streams, we often only have access to stream morphologies at present day.  
As our goal is to test the implications of time-dependence on inferred halo properties from extragalactic stream imaging, we are interested in how well the \texttt{X-Stream} sampler recovers the true present-day m12m halo parameters. 
We therefore first need a representation of the m12m $z=0$ snapshot to use as our ground truth in which we can evolve the streams from their present-day projected locations on the sky. 

We model the stellar, gas, and dark matter distributions separately. 
For each particle type, we first rotate the particle positions and velocities into the principal axis frame of the host galaxy, as determined from the inertia tensor of the stellar component.
We neglect the gas component in the potential fit because, around the location of the disk, the gas represents $<20$\% of the mass. 

For the stellar component, we assume axisymmetry and bin the star particles in a 2D cylindrical $(R,z)$ grid with $R\in(0, 40)~{\rm kpc}$ and $z\in(-20, 20)~{\rm kpc}$ with bin size $\approx 80~{\rm pc}$ in both dimensions.
We integrate the resulting density distribution over $z$ to obtain the azimuthally-averaged surface-mass density, $\Sigma_\star(R)$. 
We then fit this surface-density profile with a combination of disk and spheroid components:
We model the disk and any potential bulge component with a sum of two triple Miyamoto-Nagai (MN3) disks \citep[with $\mathrm{sech}^2$ vertical density profiles;][]{Smith:2015, Miyamoto1975} and two spherical power-law components with exponential cutoffs. 
The disk scale heights were fixed to $0.5$ and $2.0$ kpc after visual inspection of the vertical density profile, but the component masses, disk radial scale lengths, spheroid scale radii, and power-law slopes were all allowed to vary. 
We fit the profile by minimizing the logarithmic surface-density residuals,
\begin{equation}
    \mathcal{L}_\star =
    \sum_i
    \left[
        \ln \Sigma_{\star,\mathrm{model}}(R_i)
        -
        \ln \Sigma_{\star,\mathrm{sim}}(R_i)
    \right]^2,
\end{equation}
together with a constraint that the summed mass of the analytic components reproduce the total stellar particle mass. 

For the dark-matter component, we measured the spherically averaged density profile in logarithmically spaced radial shells between $r\in(0.1, 300)~\mathrm{kpc}$. 
We model this distribution with a spherical Zhao profile \citep{Zhao1996}.
We fix $\alpha=1$ 
and fit the density normalization $\rho_0$, scale radius $r_{\rm s}$, inner density slope $\gamma$, and outer density slope $\beta$ by minimizing the logarithmic density residuals,
\begin{equation}
    \mathcal{L}_{\rm DM} =
    \sum_i
    \left[
        \ln \rho_{\rm DM,model}(r_i)
        -
        \ln \rho_{\rm DM,sim}(r_i)
    \right]^2 \quad .
\end{equation}
We additionally constrain the dark-matter mass within $371\,{\rm kpc}$ to be $1.47\times10^{12}~\mathrm{M}_\odot$, based on the known virial mass of m12m, after subtracting the baryonic contribution.

For both components, we obtained optimized parameter estimates using the \texttt{Adam} optimizer \citep{kingma:2014} as implemented in \texttt{numpyro} \citep{phan2019composable, bingham2019pyro}.
We show the comparison between our Zhao fit to the dark matter particles (blue) and the FIRE particle density (black) in Figure~\ref{fig:zhao_z0}. 
We show the resulting surface-mass density profile of the fitted model of the star particles in Appendix Figure~\ref{fig:baryons-fitted}.
We list the best fit disk, bulge, and halo parameters in Table \ref{tab:BFEfit}. 

\subsection{The X-Stream sampler}\label{sec:xstream}
\citet{NibPear2025} developed the generative model \texttt{X-Stream}, which translates the projected track of an external stream into constraints on the progenitor properties (mass, integration time, orbit) and host dark matter halo properties (mass and radial profile).  They found that it was possible to recover the radial profile of the halo, and in some cases, mass limits. They found that certain streams carried more or less information about their host halo, depending on the stream's properties (e.g., length and curvature) and where they reside in the halo. 

We apply \texttt{X-Stream} to the three m12m FIRE streams. For the m12m streams we know their true present-day 6D phase-space location, and the halo properties at present day. We can thus test how well our method would work on observed streams in realistic cosmological halos.

\begin{deluxetable}{lcc}
\tablecaption{Fits to dark matter and baryonic particles  in the m12m  $z=0$ snapshot}
\tablecolumns{3}
\tablenum{2}\label{tab:BFEfit}
\tablewidth{0pt}
\tablehead{\colhead{\bf } & 
\colhead{\bf  }   &{\bf unit  }    
}
\startdata
\hline
\bf{Zhao Halo Fit}& &\\
\hline
log10($M_{\rm halo}/M_{\odot})$ & 11.1 & \\ 
$r_s$ & 2.96 & [kpc] \\ 
$\gamma$ & 0 & \\ 
$\beta$  & 2.59 & \\ 
\hline
\bf{Spheroid1}& &\\
\hline
log10($M/M_{\odot})$ & 10.1 & \\ 
$\alpha$ & 0.363 & \\ 
$c$  & 1.05 & \\ 
\hline
\bf{Spheroid2}& &\\
\hline
 log10($M/M_{\odot})$ & 10.4 & \\ 
$\alpha$ &  0.623 & \\  
$c$  & 2.36 & \\  
\hline
\bf{MN3 Disk1} & &\\
\hline
log10($M/M_{\odot})$& 10.4 & \\ 
h$_{R}$ & 1.93 & [kpc]\\ 
h$_{z}$  & 0.50 & [kpc]\\
\hline
\bf{MN3 Disk2} & &\\
\hline
log10($M/M_{\odot})$&10.9 & \\ 
h$_{R}$ & 4.14 & [kpc]\\ 
h$_{z}$  & 2.0 &[kpc] \\
\hline
\enddata
\tablenotetext{}{The baryons are fixed throughout the paper, but the halo potential is allowed to vary. The fit to the halo parameters is used as a ground truth comparison. }
\end{deluxetable}

\begin{deluxetable*}{lcccc}
\tablecaption{\texttt{X-Stream} parameters}
\tablecolumns{4}
\tablenum{3}\label{tab:freeparams}
\tablewidth{0pt}
\tablehead{\colhead{\bf } & 
\colhead{\bf Massive }  &{\bf S-shaped  } & {\bf Curlicue  } &{\bf unit  }    
}
\startdata
\hline
\bf{Free params (vrad free)}& &&\\
\hline
$y_{\rm prog}^*$ &   [-130, 130]&  [-130, 130]&[-130, 130]& [kpc]   \\
$\eta_r$  & [-1, 1] &  [-1, 1] &[-1, 1]  &\\
$\eta_t$  & [0, 1.5] & [0, 1.5]& [0, 1.5]&  \\
$\psi$&[0, 2$\pi$]& [0, 2$\pi$]&[0, 2$\pi$]&\\
log$_{10}m_{\rm prog}$$^*$&[7.5, 10.0]&[7.5, 10.0]&[7.5, 10.0]&[M$_{\odot}$]\\
log$_{10}m_{\rm halo}$$^*$&[10, 13]&[10, 13]&[10, 13]& [M$_{\odot}$]\\
$r_s^*$   &  [2, 20]  & [2, 20]&[2, 20]&[kpc]  \\
$\gamma^*$& [0, 2]&[0, 2]&[0, 2]\\
$\beta^*$ &[2, 4]&[2, 4]&[2, 4]\\
$t_{\rm age}^*$ & [0.5,5]& [0.5, 3.5]&[2, 6]&[Gyr]\\
\hline
\bf{Fixed params (vrad free)}& &&\\
\hline
$x_{\rm prog}^*$ & 62.1 &$-68.3$ & $-11.9$& [kpc]  \\
$z_{\rm prog}^*$ & 66.4&24.6& 72.9& [kpc]   \\
\hline
\bf{Different free params (vrad fixed)}& &&\\
\hline
$v_x$  & [-0.6, 0.6] &  [-0.6, 0.6] &[-0.6, 0.6]  & [kpc/Myr]\\
$v_z$  & [-0.6, 0.6] & [-0.6, 0.6]& [-0.6, 0.6]& [kpc/Myr] \\
\hline
\bf{Additional fixed params (vrad fixed)} & &&\\
\hline
$v_{y,\rm los}$ &124& 74.5 & 116&[km/s] 
\enddata
\tablenotetext{}{$^*$ These parameters are the same for all runs. }
\end{deluxetable*} 

Throughout this paper, we use two different sampling approaches for the velocities of the streams. The first approach assumes no kinematic information of the stream is available, while the second assumes that the progenitor's radial velocity has been measured.
In the first approach,  we  define $v_{\rm circ}$ as the local circular velocity at the progenitor's position. 
We sample two velocity components: along the radial direction, $v_r$, and along the 
tangential direction, $v_t$, expressed as fractions of $v_{\rm circ}$:
\begin{equation}
    \eta_r = \frac{v_r}{v_{\rm circ}} \qquad \text{and} \qquad \eta_t = \frac{v_t}{v_{\rm circ}}.
\end{equation}
We sample these parameters uniformly: $\eta_r \in [-1, 1]$ and $\eta_t \in [0, 1.5]$, favouring tangentially supported orbits over radially supported ones, consistent with coherent streams rather than shells. We additionally 
sample a velocity angle $\psi \in [0, 2\pi]$, giving the velocity vector
\begin{equation}
    \mathbf{v} = \eta_r v_{\rm circ}\,\hat{r} 
               + \eta_t v_{\rm circ} \left[\cos\psi\,\hat{u}_1 + \sin\psi\,\hat{u}_2\right],
\end{equation}
where $\hat{u}_i$ are unit vectors perpendicular to $\hat{r}$ and to each other, defined as: 
\begin{equation}
    \hat{u}_1 \equiv \frac{\hat{r} \times \hat{z}}{\Vert \hat{r} \times \hat{z} \Vert} \qquad \text{and} \qquad 
    \hat{u}_2 \equiv \hat{r} \times \hat{u}_1.
\end{equation}
Sampling $\psi \in [0, 2\pi]$ allows the stream to circulate in either direction; 
the circulation direction is encoded in $\psi$, which is why $\eta_t \geq 0$.

\begin{figure*}
    \centering
    \includegraphics[width=\textwidth]{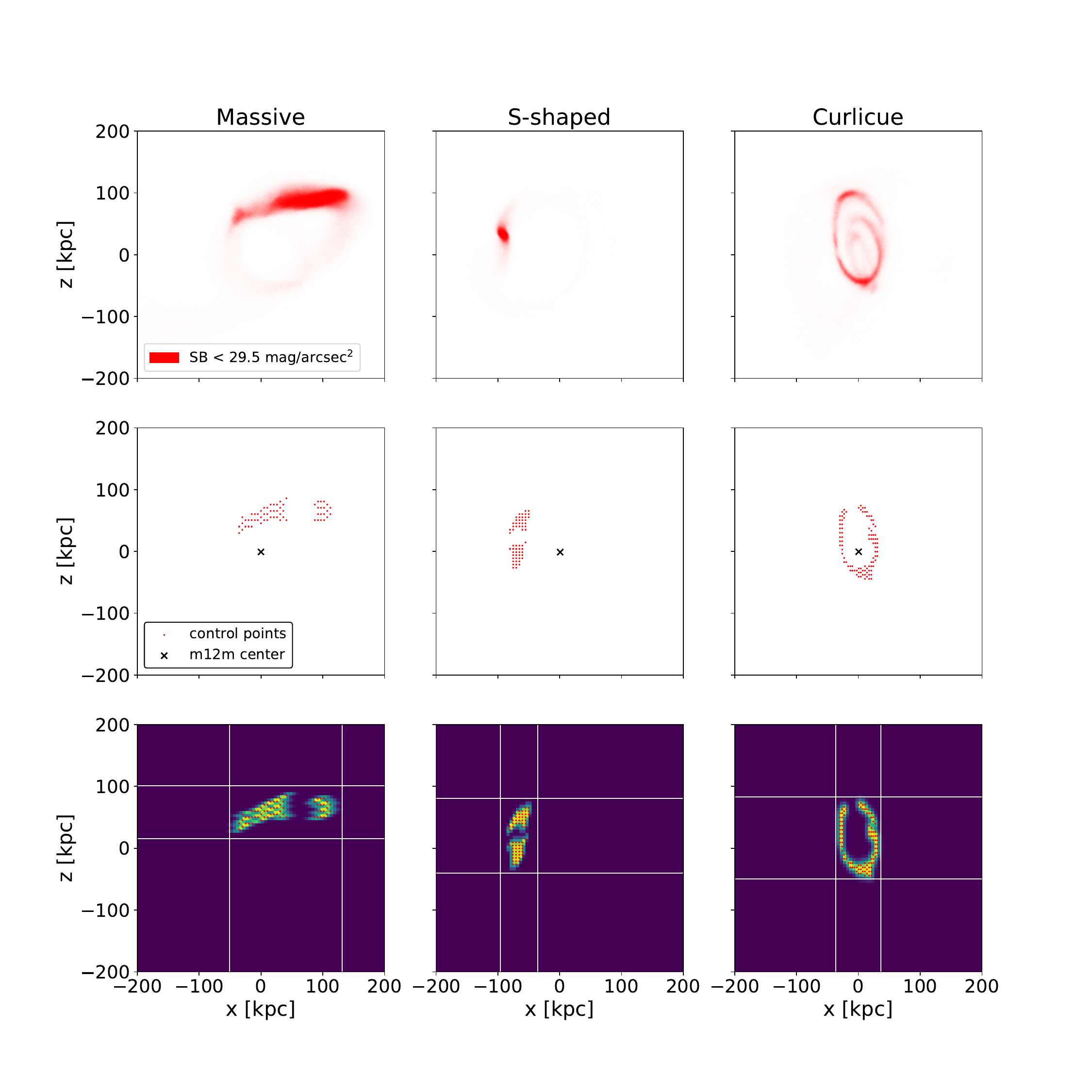}
    \caption{{\bf Top}: 
    All star particles which are brighter than the wide field {\it Euclid} survey limits (SB$<$ 29.5) from Figure \ref{fig:sim} for each of the three streams at $z=0$. Here $x,z$ defines the plane of the sky. Darker red regions show a higher density of stars.  
     {\bf Middle}: Control points (red) used as input to \texttt{X-Stream}. The top-row pixels are binned into 1 kpc bins and the bins in the top 10\% by particle count are retained, then subsampled onto a uniform grid with 5 kpc spacing for the Massive and S-shaped streams and 2.5 kpc spacing for the Curlicue stream. The red points mark the centers of the retained bins.
    We have masked out red control points at the progenitor regions and the disk of m12m. The black x marks the center of m12m in each panel. 
    {\bf Bottom}: KDE from control points and bounding boxes used in \texttt{X-Stream}, with a bandwidth of 0.15 kpc. We also mask out the progenitor regions here since particle-spray models do not well reproduce the escape conditions of  dwarf galaxies. 
    }
    \label{fig:projections}
\end{figure*}

In our second approach,  we fix the progenitor's radial velocity to the true value from m12m and sample uniformly over the two other velocity components. 
These two approaches are different from the \citet{NibPear2025} method. In \citet{NibPear2025}  they uniformly sampled the velocity unit sphere, which tends to oversample radially plunging orbits. 

To apply the \texttt{X-Stream} sampler, we first need to define control points tracing the morphology of each stream to be used as input data. 
In the top row of Figure  \ref{fig:projections}, we show all stream particles which belong to the three streams in m12m from Figure \ref{fig:sim} and which had star particles brighter than 29.5 $\rm{mag} \, \rm{arcsec}^{-2}$ as per the {\it Euclid} wide field limits. We ignore the stellar halo and disk stars in m12m and use the stream particles tagged by \citet{Panithanpaisal2021,Horta2023} directly.

Because most external streams in {\it Euclid} data will not have resolved stars, we use a binning procedure to construct the mock input for the \texttt{X-Stream} code. We bin the particle data in the $x-z$ plane into $1~\mathrm{kpc}^2$ bins, retaining only the top 10\% of bins by particle count.
Bins in the top 10\% are given a value of 1, while all others are given a value of 0. The result is typically a very jagged sampling of the stream's 2D track and width. To smooth out the streams for our analysis, we subsample the surviving bins onto a uniform grid with 5 kpc spacing for the Massive and S-shaped streams, and 2.5 kpc spacing for the Curlicue stream,  since it is a thinner stream. The retained bin centers are the control points used as input to \texttt{X-Stream} (see middle panels of Figure \ref{fig:projections}). In Section \ref{sec:extendeddebrisdiscus}, we discuss how our results change if we instead keep the top 30\% of bins.

In the bottom row of Figure  \ref{fig:projections} we show a  kernel density estimate (KDE) of each stream generated from the red control points  \citep[see details in][]{NibPear2025}. The KDE has a bandwidth of 0.15 kpc. 
We mask out a radius of 25 kpc around the Massive stream progenitor and 12 kpc around the S-shaped and Curlicue progenitors as the particle-spray method is not designed to capture the details of the escape conditions near the progenitors. We also mask out the m12m disk region for the Curlicue stream, as  stream stars in this region would be difficult to disentangle from the disk. 

We generate model stellar streams in the analytic representation (see Section \ref{sec:potential}) of m12m's potential at $z=0$, 
using the particle-spray technique by \citet{Fardal2015} implemented in the GPU accelerated code \texttt{streamsculptor} \citep{Nibauer2025a}, which utilizes the \texttt{Jax} Python framework \citep{Jax2018}. We fix the escape conditions to those of \citet{Fardal2015} (see their  $k$-values). 
In our simulations ($x, z$) defines the sky plane and $y$ is in the line-of-sight direction. 
We run the simulations on an NVIDIA A100 GPU. 

\texttt{X-Stream} generates stream models in batches of 200, each in a different potential, to identify the progenitor and halo properties that reproduce the FIRE stream morphologies. 
At every choice of model parameter (see priors in Table \ref{tab:freeparams}), we generate a mock-stream. We do not penalize models that are longer than the observed stream, as particles beyond the bounding boxes in the bottom row of Figure \ref{fig:projections} are ignored. We fit each model with a 2D KDE (bandwidth = 0.15 kpc) in the same way as for the input data described above. 

To determine how well our generated model streams fit the representation of the FIRE m12m streams at $z=0$, we use  KL divergence \citep{kullback1951} statistic between the model stream KDE and the mock-observation control points (red points, Figure~\ref{fig:projections}). Nested sampling \citep{nautilus} is utilized to build up an approximate posterior distribution. Further details of the sampling procedure are provided in \citet{NibPear2025}.

Throughout the paper, we fix the disk and bulge parameters from the $z=0$ fit in Section \ref{sec:FIRE}, as we assume that the disk and bulge properties can be estimated from observations, while the halo parameters are free. We also fix the present-day $x-z$ location of the three stream progenitors, while keeping the line-of-sight position, mass, velocities, and integration time free. 
In observations, we most often do not know the progenitor location, and future work could relax this assumption.
We use the same range of priors for the free parameters for the three streams except for integration time, 
where we allow for longer integration times, for the extended debris runs and longer streams (see Table \ref{tab:freeparams} and Appendix \ref{sec:corner}). 

Our sampler rejects unbound progenitor orbits. Similarly to the definition in \citet{pearson2024} of how ``stream-y'' tidal debris appears, we also reject stream models for which the stars are dispersed symmetrically around the center of the galaxy in the plane of the sky (in this work: $x,z$). Specifically, we reject models where $\Delta_0 \equiv \sqrt{{\rm med}(x)^2  + {\rm med}(z)^2} < 10$ kpc (where med denotes median). Stream models with small values of $\Delta_0$
represent fully phase-mixed streams or streams near the center of the galaxy, obscured by the disk.

The particle-spray method strips stars uniformly in time, including orbital phases where the tidal radius far exceeds the progenitor size and stripping should not occur.
Similarly to \citet{pearson2022b}, we, therefore, apply a Jacobi radius condition: stars only strip when $r_J < 3\times r_e$. Orbits for which this condition is never satisfied are rejected. 
We choose a factor of 3 to avoid over-rejecting orbits. Lower thresholds would reject too many, while larger thresholds would allow for stripping throughout the entirety of the orbits. 
We strip stars only if the condition is met.
This encourages stripping primarily near pericenter, as expected for dwarf galaxy disruption \citep[e.g.,][]{bonaca2025}, and is an update to the \citet{NibPear2025} method. 
To define a progenitor size for each model stream generated with \texttt{X-Stream}, we fix the present-day $r_e$ of each model stream progenitor from a power-law extension of the \citet{shen2003} SDSS size-mass relationship, which was used in \citet{prole2021} (Eq. 6).

Observations of radial velocities have been measured for some extragalactic streams \citep[e.g. M31,  Centaurus A, NGC 4449][]{Gilbert2009,denja2016,Toloba2016}. In other cases, globular clusters can serve as kinematic stream tracers if their projected locations coincide with stream features \citep{hughes2023,mueller2025}, and planetary nebulae radial velocity measurements along streams will become readily available in the coming years \citep{Valenzuela2026}.  We therefore run \texttt{X-Stream} using both fixed and free line-of-sight velocities for the progenitor (see Table \ref{tab:freeparams}).

\section{Results} \label{sec:Results}
In this Section, we first test whether the static potential (Section \ref{sec:potential}) is capable of reproducing the mock observations using the true 6D phase space coordinates of each stream  (Section \ref{sec:nofit}).
We then apply \texttt{X-Stream} to fit the present-day morphology of the FIRE m12m streams and test how well we recover the halo and progenitor parameters under the assumption of a static halo (Section \ref{sec:fitting}). 
Since six corner plots with 10 and 9 parameters are hard to digest at once,  we  first show results from the orbital constraints in Section \ref{sec:orbit}, then radial density constraints in Section \ref{sec:density}, and lastly we show the halo and progenitor mass constraints in Section \ref{sec:mass}. 

\subsection{True orbits and the tidal radius condition}\label{sec:nofit}
For each of the three stream progenitors in m12m, we know the present-day 6D phase-space position from the 500 particles closest to the highest density location in the unwrapped streams in the simulation.
To test how well the particle-spray model and a static potential reproduce these mock observations, we first integrate each progenitor’s orbit backwards for 5~Gyr in the fitted m12m $z=0$ potential (Section~\ref{sec:methods}). We then evolve each stream forward using particle-spray, releasing two star particles every 1 Myr.
 
This test ignores any disk and halo growth, any change to the progenitor orbits due to dynamical friction or halo mass growth, and any evolution in halo shape. 
We carry out this test with and without the tidal radius condition introduced in Section  \ref{sec:xstream}, and fix the progenitor masses to the total stellar masses of the streams and intact progenitors at $z=0$ (see Table \ref{tab:streams}). We find evidence for perturbations between two of the progenitors which we discuss later in this section. 

\begin{figure*}
    \centering
    \includegraphics[width=0.8\textwidth]{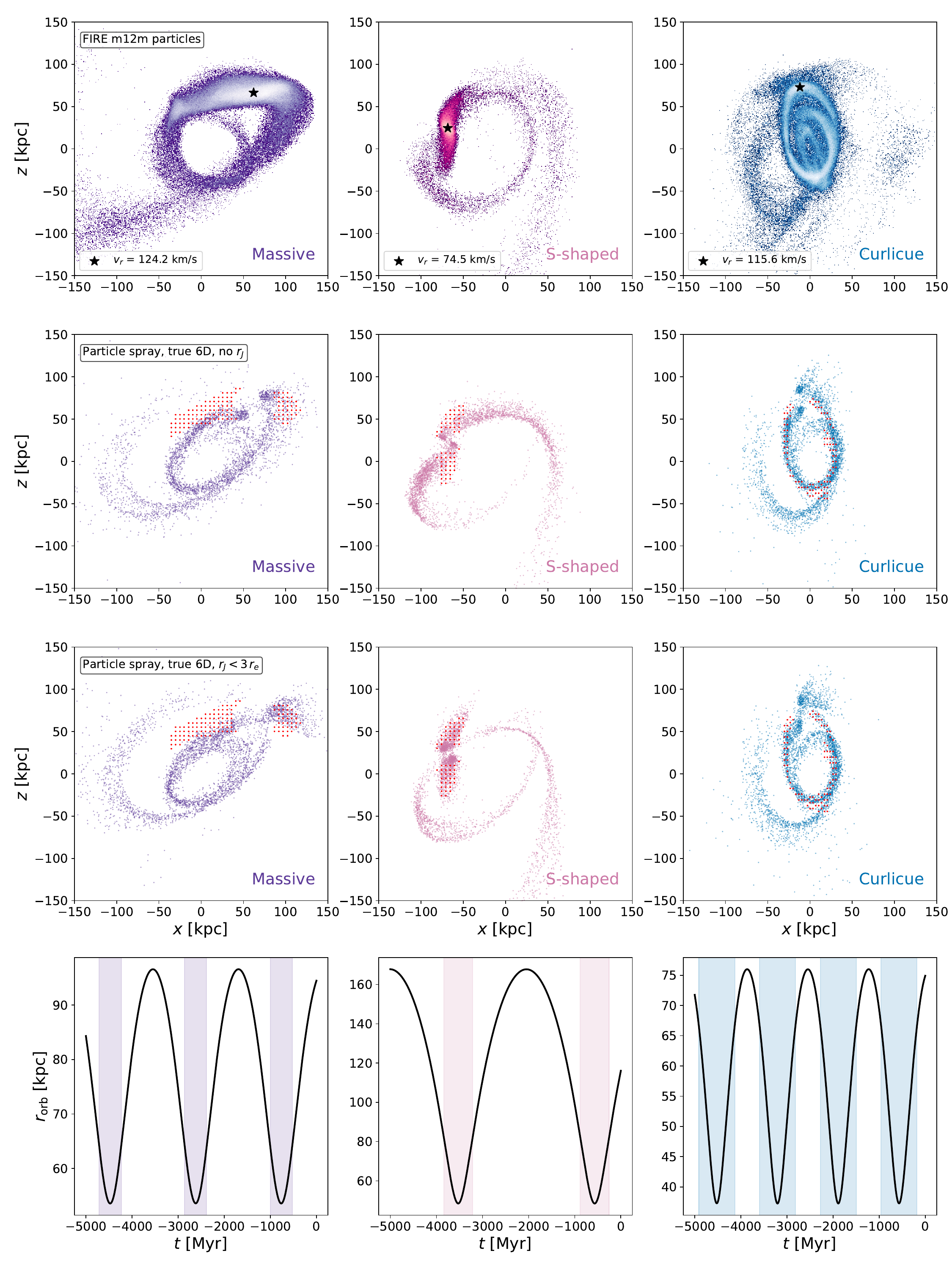} 
    \caption{
    {\bf 1st row}: All FIRE m12m particles belonging to each stream as tagged by \citet{Panithanpaisal2021} and \citet{Horta2023}.
    The black stars indicate the progenitor position, and the label states each progenitor's radial velocity at present day. 
    {\bf 2nd row}:  particle-spray streams simulations integrated for 5 Gyr from the true present-day 6D phase space information (see Table \ref{tab:streams}) in static halo representation of the $z=0$ snapshot from the m12m particle data. The red control points are the same as shown in Figure \ref{fig:projections} for reference. No fitting was involved in generating these streams, and stars were stripped uniformly throughout the entire orbital history. 
    {\bf 3rd row}: Same as 2nd row but now applying a tidal radius condition where stars only strip if $r_J< 3 \times r_e$ (see shaded regions in the bottom row). 
    Applying the $r_J$ condition changes the apparent morphologies of the Massive and S-shaped streams, and the S-shaped and Curlicue streams (3rd row) look remarkably similar to the FIRE streams (top row) despite our assumptions of static halo.
    {\bf 4th row}: Galactocentric radius of each progenitor as a function of time (black line) for the true 6D orbits evolved in a static potential.  Present day ($z=0$) is at $t=0$. We visualize the cosmological orbits from FIRE m12m for each progenitor in Appendix Figure \ref{fig:infall}. }
    \label{fig:true6D}
\end{figure*}

In the top row of Figure \ref{fig:true6D} we show all particles that belong to each of the three streams in FIRE m12m tagged by \citet{Panithanpaisal2021} and \citet{Horta2023}. The black star indicates the location of the progenitor.
In the second row of Figure \ref{fig:true6D} we show the result of our experiment for streams generated with the particle-spray technique. In this row, particles are stripped uniformly in time without any tidal radius condition. The red control points trace the densest parts of the FIRE stream stars brighter than the {\it Euclid} survey limits (29.5 $\rm{mag} \, \rm{arcsec}^{-2}$) as described in Figure \ref{fig:projections}.
The particle-spray model of the Massive stream (left) is more tightly wound, with a higher curvature than the present-day streams in FIRE m12m (top). The left arm is offset from the red control points, and the diffuse extended arm to the lower left is also more tightly wound in the particle-spray model.   
The S-shaped model stream (middle) has a similar curvature to the red control points, but its angle is offset from the red control points. 
The Curlicue stream (right) looks  similar to the FIRE stream (top) but with less dispersed debris. 
Note that the dense blobs of material near the progenitor locations shows the most recently stripped material which has not yet phase mixed. %

In the third row of Figure \ref{fig:true6D} we show the same particle-spray simulations but after applying the tidal radius condition, where stars are only allowed to strip if the tidal radius is smaller than three times the extent of the initial dwarf progenitor (see Section \ref{sec:methods}). 
In the bottom row, we show the galactocentric distance over time for each progenitor orbit evolved in the static halo. The shaded areas highlight where the tidal radius condition is fulfilled and stars can strip. For the Massive and S-shaped streams stripping only occurs around pericenter, where for the Curlicue stream, only stars stripped very close to apocenter are omitted. 

After applying the tidal radius condition, the Massive stream still appears offset from the control points (3rd row, left). Several time-dependent effects in  m12m are not captured by our static model. 
In the m12m evolution, the leading arm of the Massive stream appears visually perturbed by an encounter with the S-shaped progenitor at $t\sim -0.8$ Gyr.  We do not quantify the encounter strength, but at closest approach the S-shaped progenitor comes within 1 kpc of the Massive stream's leading arm, with a relative velocity $<20$ km/s. This encounter likely explains the offset in this stream \citep[see e.g., LMC-perturbed streams in][]{shipp2021}. 
Separately, Appendix Figure \ref{fig:infall} shows that the Massive stream progenitor's orbit decays measurably  over the last 6 Gyr, unlike the S-shaped and Curlicue progenitors, whose orbits show no decay over the same period (the Curlicue progenitor fully disrupts at $t\sim -3.8$ Gyr). The present-day stellar mass of each progenitor (if intact) plus its stream (Table \ref{tab:streams}), relative to the m12m host halo mass, is $\sim$0.070\%, $\sim$0.022\%, and $\sim$0.039\% for the Massive, S-shaped, and Curlicue streams, respectively. For massive streams, dynamical friction and more complicated stripping from the progenitor is hard to model in our current framework. The orbital decay could also be affected by the halo's mass growth or by adiabatic contraction in the inner regions. 
While we do not include encounters with perturbers or orbital decay in our stream modeling directly, we can test how the bias propagates into the \texttt{X-Stream} inference, and observationally, we would be able to distinguish massive progenitors from less massive systems from surface brightnesses of streams. 
Both the Curlicue and S-shaped streams (3rd row) resemble the FIRE m12m streams (top row) after applying the tidal radius condition, and the recently stripped material for both trace the  red control points well despite being evolved in a static halo. 
The extended debris beyond the red control points is slightly offset in angle for the S-shaped stream. This is consistent with findings of \citet{buist2015}, who showed that halo mass growth does not significantly affect the morphologies of recently released stream stars, but changes the angles of previously stripped debris.

Despite the Massive stream mismatch, the S-shaped and Curlicue streams are promising: their recently stripped material appears insensitive to halo time-dependence, and they are good candidates for the {\texttt X-Stream} sampler based on  intuition from \citet{NibPear2025}.

\subsection{Applying X-Stream}\label{sec:fitting}

To test whether we can learn about true halo properties, stream progenitors, and stream orbits from extragalactic stellar streams under the assumption of a static potential, we run \texttt{X-Stream} separately on all three streams. First we apply \texttt{X-Stream} with 10 free parameters using the red control points  described in Figure \ref{fig:projections} as input data, and then with 9 free parameters, where we fix the radial velocity of the progenitor. 
We list the range of priors for the runs in  Table \ref{tab:freeparams}.

\begin{figure*}
    \centering
    \includegraphics[width=\textwidth]{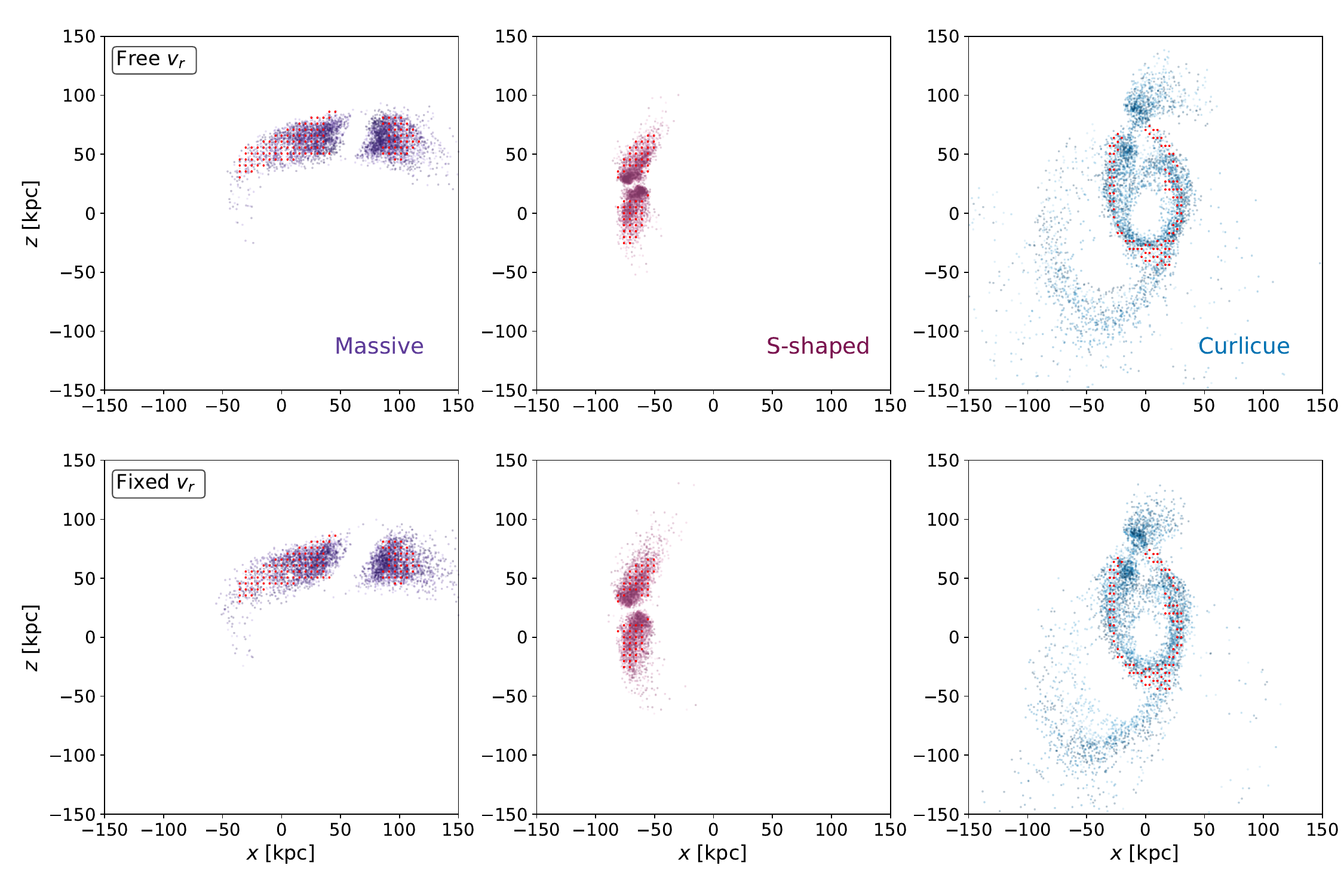}
    \caption{Examples of streams randomly selected within the 68\% posterior confidence region for the Massive stream (left), S-shaped stream (middle), and Curlicue stream (right) run in \texttt{X-Stream} with 10 free parameters (top) and while fixing the radial velocity and using 9 free parameters (bottom). Each panel show examples of five different overplotted streams. 
   All parameters for each stream in this Figure are shown as  white points in the full corner plots presented in Appendix \ref{sec:corner}. All streams reproduce the input data (red control points) well. }
    \label{fig:truevsfit}
\end{figure*}

We show the posterior distributions for each of the six \texttt{X-Stream} runs  in Appendix \ref{sec:corner}. In each corner plot we show five randomly selected points which fall within the 68\% confidence regions of the posterior distributions. The black lines show the truth values from our fitted potential at $z=0$ and from our determination of the true progenitor 6D location at $z=0$.

In Figure \ref{fig:truevsfit}, each panel shows the five randomly selected streams (overplotted) from the 68\% confidence region of the posterior distributions. 
Each panel shows the red control points used as input for \texttt{X-Stream}. 
The top panel shows our inference with 10 free parameters and the bottom row show inference with a fixed progenitor radial velocity.

For the Massive stream (Figure \ref{fig:truevsfit}, left), the five random model streams from within  the 68\% confidence regions of the posterior distributions  trace the red control points well.  When using the true present-day phase-space location of the progenitor under our static potential and particle-spray assumptions (Section \ref{sec:nofit}), the model failed to reproduce the stream; however, our inference is capable of finding parameters that successfully matches its morphology. This is the case both for the run with 10 free parameters (top), and the run where the radial velocity was fixed (bottom). 
Bias in model parameters will be discussed in the subsequent sections. 

The S-shaped streams (Figure \ref{fig:truevsfit}, middle) look similar to the most recently stripped debris of the S-shaped FIRE stream in both the top (free radial velocity) and bottom (fixed radial velocity) case, and  trace the red control points well. 
In Section \ref{sec:extendeddebrisdiscus}, we discuss \texttt{X-Stream} runs, where we include more extended control points as input data for the massive and S-shaped streams.

The Curlicue streams (Figure \ref{fig:truevsfit}, right) trace the red control points, but also reproduce the extended wrap of the stream. This wrap looks similar to the true stream in FIRE m12m (see Figure \ref{fig:true6D} top row), despite the red control points not tracing this region both for the run with 10 free parameters (top), and the run where the radial velocity was fixed (bottom).

In summary,  \texttt{X-Stream}  reproduces the morphology of the red control points well for each stream, as expected from \citet{NibPear2025}, and there are no noticeable morphological differences in the streams from the runs with free radial velocity (top) versus the fixed radial velocity runs (bottom).
A good visual fit to stream morphology does not necessarily equate to correct inference on orbital and potential parameters. We explore constraints on these parameters in the next section.

\subsubsection{Orbital constraints}\label{sec:orbit}
For simplicity, throughout the remainder of this Section we present constraints from the \texttt{X-Stream} runs with 
free radial velocity (10 free parameters), and only comment  on differences from the runs with 9 free parameters and a fixed progenitor radial velocity. For a summary of all runs and corner plots, see Appendix \ref{sec:corner}. 

In Figure \ref{fig:orbit_massive}, we present the orbital parameter constraints for y$_{\rm prog}, \eta_r, \eta_t, \psi$  using \texttt{X-Stream} for the Massive stream. See Section \ref{sec:methods} for intuition on each of these orbital parameters. 
The purple contours show the 68\% (solid lines) and 95\% (dashed lines) confidence regions. We overplot the true 6D parameters from FIRE m12m as black lines. The white points represent the five randomly selected solutions that fall within the  68\%  confidence region, for which we visualized the respective streams in Figure \ref{fig:truevsfit} (top, left). The 1D histograms show the marginalized posterior of the full distributions.

The $y_{\rm prog}$ posterior is bimodal, with the positive mode peak slightly offset from the truth ($y_{\rm prog} = 26$ kpc). The bimodality in the $y_{\rm prog}$ posteriors reflects that we cannot determine whether the progenitor is in front of or behind the host galaxy from the morphology alone (see \citealt{pearson2022b} and \citealt{NibPear2025} for a discussion of this degeneracy). 
For the \texttt{X-Stream} run where the radial velocity is fixed, this degeneracy is broken, and there is a lower limit on $y_{\rm prog}$, but the truth is still offset from the posterior at the lower end (see Appendix Figure \ref{fig:corner_inner_vr}).  

The $\eta_r$ posterior runs into the upper prior boundary ($\eta_r = 1$) with a 68\% lower limit of $\eta_r > 0.23$. The true value, $\eta_r = 0.13$, lies outside the 68\% region but within the 95\% region ($\eta_r > -0.04$).
The tangential velocity parameter is well recovered: the true $\eta_t = 0.78$ lies within the 68\% region, with a 
posterior median and 68\% interval of $\eta_t = 0.88^{+0.17}_{-0.31}$. The true $\psi = 3.46$ lies at the lower edge of the posterior, which spans $\psi \in [3.1, 6.0]$ at 95\% confidence. For the run where we fixed the Massive stream progenitor's radial velocity, the velocity posteriors produce closed constraints, yet with the truth values at the lower edge of the 95\% interval (see Figure \ref{fig:corner_inner_vr}). 

The recovered orbital parameters for the Massive stream show mixed agreement with the true parameters.  
This  was  expected from our discussion of the particle-spray simulation of the Massive stream in Section \ref{sec:nofit} and Figure \ref{fig:true6D}, and since the Massive stream experiences significant orbital decay over the past 6 Gyr (see Appendix Figure \ref{fig:infall}). 
\texttt{X-Stream} uses only the present-day morphology of the stream as input. If the leading arm has been perturbed by an encounter (see Section \ref{sec:nofit}) its morphology no longer traces the progenitor's orbit, and we would expect biased orbital parameters when comparing to the true 6D values.
The red control points have lower curvature than the stream evolved from the true 6D parameters in a static halo (Figure \ref{fig:true6D}), which explains why the posteriors favour orbits with larger total velocity.

\begin{figure}
    \centering
    \includegraphics[width=\columnwidth]{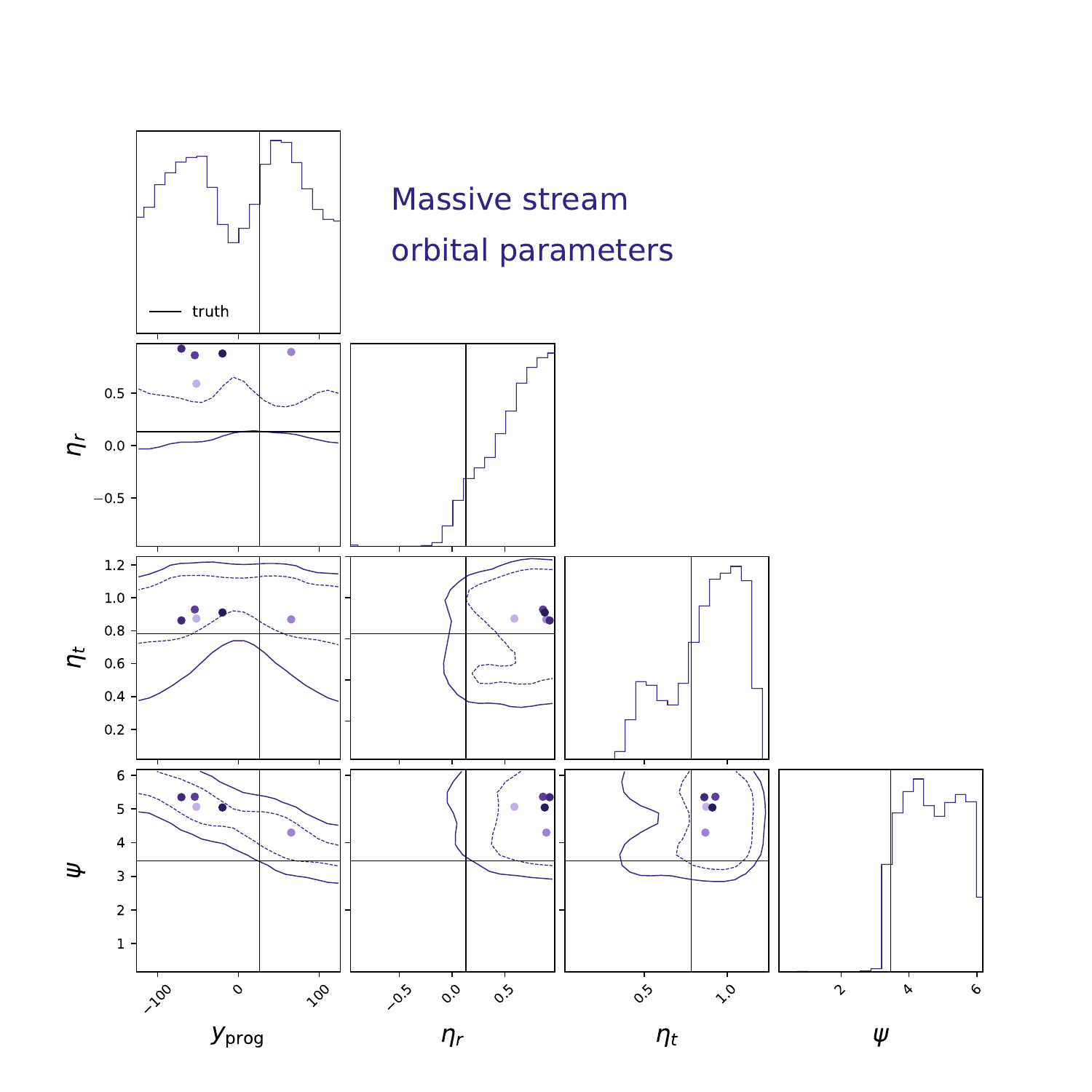}
    \caption{ Summary of orbital constraints for the Massive stream. Purple dashed contours show the 68\% levels, the solid purple lines show the 95\% levels in both the contours and 1D histograms. The black lines show the true values extracted from the m12m simulations. 
 } 
    \label{fig:orbit_massive}
\end{figure}

\begin{figure}
    \centering
    \includegraphics[width=\columnwidth]{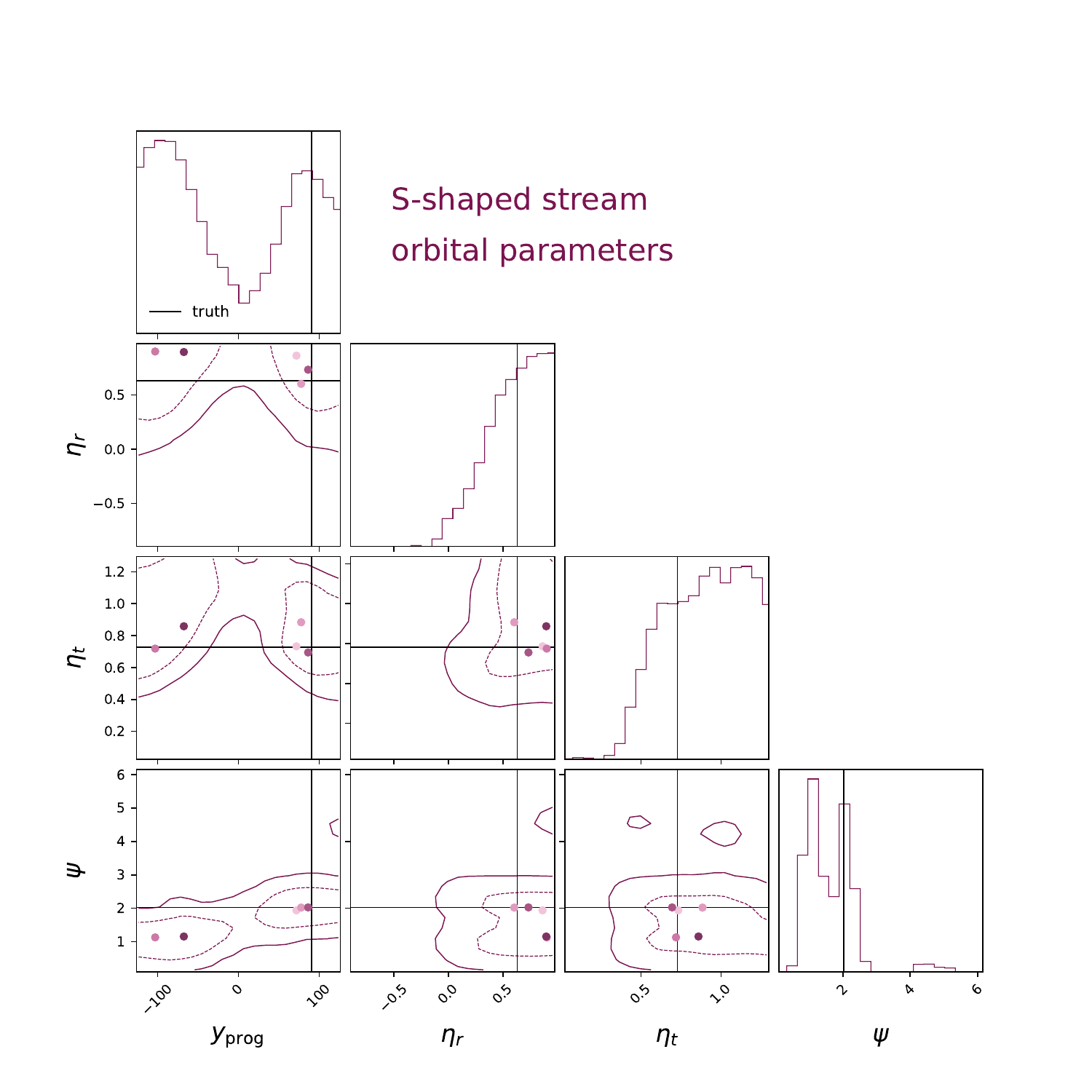}
    \caption{Summary of orbital constraints for the S-shaped stream. Pink dashed contours show the 68\% levels, the solid pink lines show the 95\% levels in both the contours and 1D histograms. The black lines show the true values extracted from the m12m simulations. 
 } 
    \label{fig:orbit_sshaped}
\end{figure}

In Figure \ref{fig:orbit_sshaped}, we present the orbital parameter constraints for the S-shaped stream. All four true orbital parameters lie within the 68\% confidence region. The $y_{\rm prog}$
posterior is bimodal, with the true value ($y_{\rm prog} = 90.6$ kpc) lying within the positive mode. 
For S-shaped stream, fixing the progenitor's radial velocity in the fits does not break the $y_{\rm prog}$ degeneracy (see Figure \ref{fig:corner_outer_vr}). The joint posterior in $t_{\rm age}$ and $y_{\rm prog}$ shows that viable solutions in the negative (incorrect) $y_{\rm prog}$ mode are restricted to short integration times, with high $M_{\rm halo}$, $M_{\rm prog}$, and velocities. 
We have checked that the short integration time does not give the progenitor enough time to reach positions where the acceleration field differs enough to produce distinguishable morphologies, leaving the 
$y_{\rm prog}$ degeneracy unbroken even with the radial velocity fixed. Additionally, the tidal field in the vicinity of the progenitor at present-day is weak, further contributing to this degeneracy remaining unbroken.

As for the Massive stream, the $\eta_r$  and $\eta_t$ posteriors for the S-shaped stream rise toward their upper prior boundaries. We find 68\% lower limits of $\eta_r > 0.28$ and  $\eta_t > 0.59$ , with the true values ($\eta_r = 0.63$, $\eta_t = 0.73$) lying within the 68\% region. For $\psi$ we find a posterior median and 68\% interval of $1.33^{+0.70}_{-0.55}$, with the true $\psi = 2.03$ at the upper edge of the 68\% interval. When we fix the radial velocity of the S-shaped progenitor,  the true orbital values are also within the 68\% region,  but the posteriors still only place lower limits on velocity (see Figure \ref{fig:corner_outer_vr}).

\begin{figure}
    \centering
    \includegraphics[width=\columnwidth]{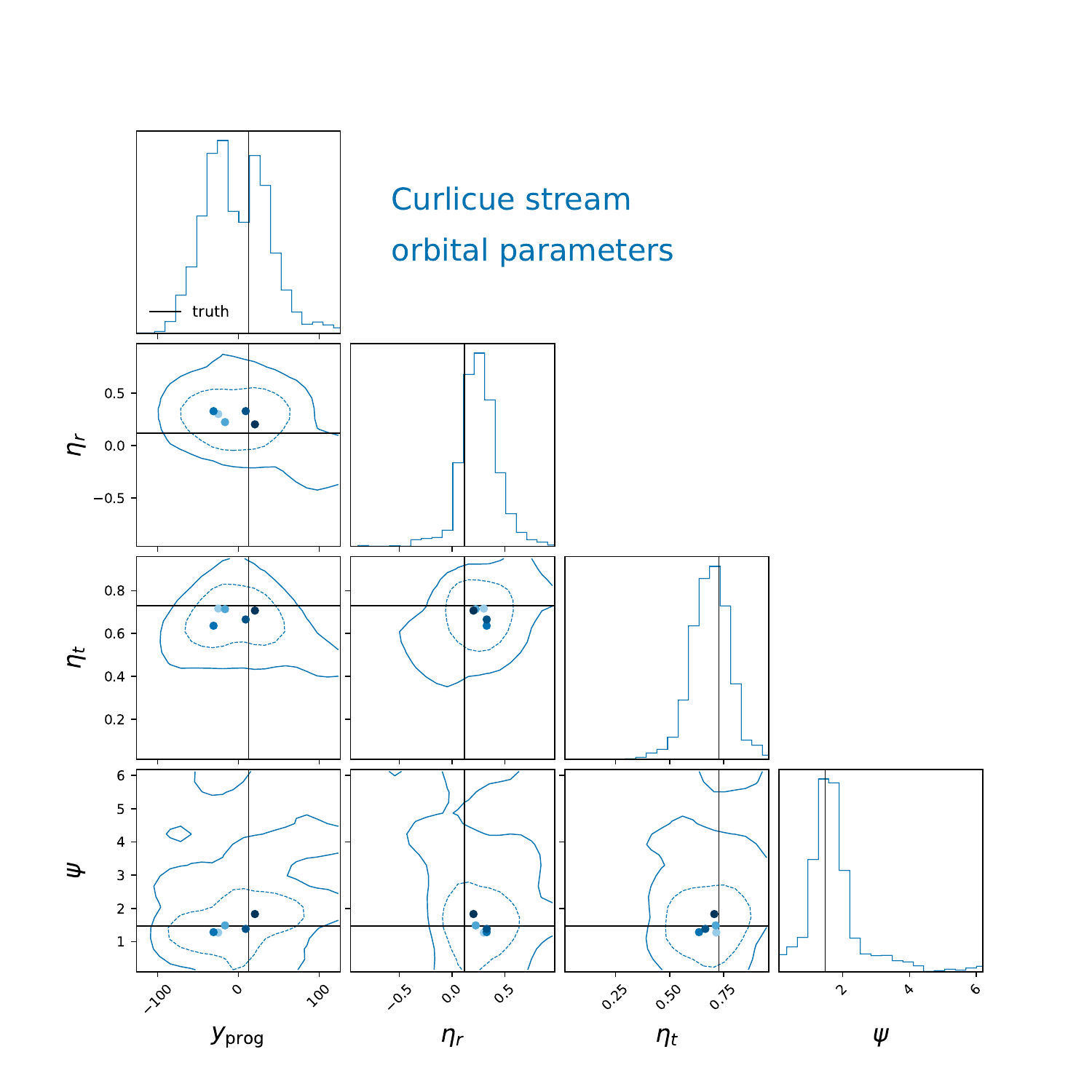}
    \caption{Summary of orbital constraints for the Curlicue stream. Blue dashed contours show the 68\% levels, the solid blue lines show the 95\% levels in both the contours and 1D histograms. The black lines show the true values extracted from the m12m simulations. All orbital parameters are well constrained, with a bimodality in $y_{\rm prog}$. 
 } 
    \label{fig:orbit_curlicue}
\end{figure}

In Figure \ref{fig:orbit_curlicue}, we present the orbital parameter constraints for the Curlicue stream. The $y_{\rm prog}$ posterior is again bimodal, with modes at $y_{\rm prog} \approx -35$ and $20$ kpc and the true value ($y_{\rm prog} = 12.6$ kpc) close to the left edge of the positive mode. The remaining posteriors peak well within the prior ranges, yielding closed constraints: posterior medians and 68\% intervals of $\eta_r = 0.20^{+0.17}_{-0.15}$, $\eta_t = 0.66^{+0.08}_{-0.09}$, and $\psi = 1.45^{+0.67}_{-0.52}$, compared to true values of $\eta_r = 0.12$, $\eta_t = 0.73$, and $\psi = 1.5$. All four true parameters are recovered within the 68\% confidence region. This is also the case for the run where we fixed the radial velocity of the Curlicue progenitor (see Figure \ref{fig:corner_curlicue_vr}). The posteriors for this run have tighter constraints, and the $y_{\rm prog}$ degeneracy is broken, with the true value recovered. 

To summarize, if we  only trace the densest parts of the streams, observable in wide field {\it Euclid} data, the Curlicue stream is the only case in which the stream morphology alone constrains the full orbit. However, limits on the progenitor orbit can be placed for shorter, less informative streams. For streams resulting from more major mergers or with recent encounters, we expect potential bias in orbital parameters at the $2\sigma$ level.

\begin{figure*}
    \centering
    \includegraphics[width=\textwidth]{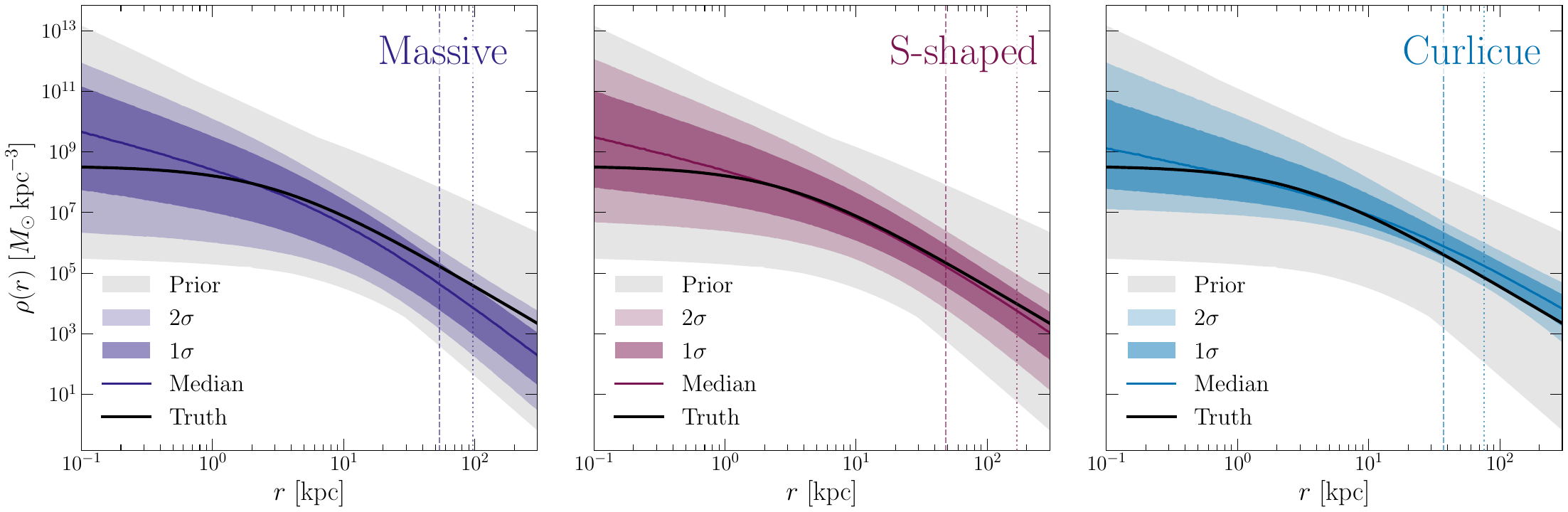}
    \caption{Constraints on  m12m's dark matter radial density profile as a function of distance from the center of the host galaxy for the massive (left), S-shaped (middle), and Curlicue (right) streams. The dark shaded regions show the 68\% confidence regions sampled over log$_{10}M_{\rm halo}$, $r_s$, $\gamma$, $\beta$,  the light regions show 95\% confidence regions, and the solid colored lines show the median of the samples. The gray area shows the prior range, and the black line shows the fitted Zhao dark matter radial density profile fit from the m12m dark matter particles at $z=0$. We do not show the baryons here. The vertical dashed and dotted lines show peri- and apocenters of each progenitor's true 6D orbits evolved in a static potential. 
    All three streams show constraints across all radii. The Curlicue stream (right) places the tightest constraints, which is also reflected in the parameter posteriors presented in Appendix \ref{sec:corner}.   }
    \label{fig:radialprofile}
\end{figure*}

\begin{figure*}
    \centering
    \includegraphics[width=\textwidth]{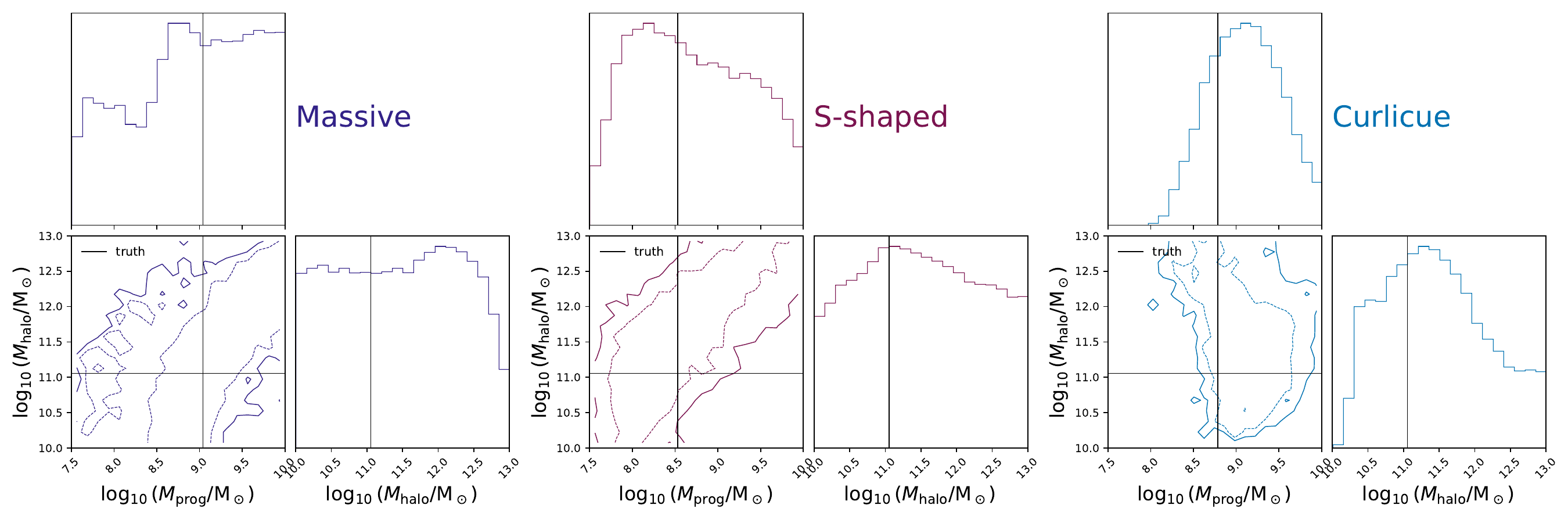}
    \caption{Summary of \texttt{X-stream} halo mass and progenitor mass constraints for the three streams. Dashed contours show the 68\% levels, the solid lines show the 95\% levels in both the contours and 1D histograms. The black lines show the true values extracted from the m12m simulations, where the halo mass is the Zhao scale mass, and the progenitor mass is defined as the total stellar mass in the stream and progenitor at present day. }
    \label{fig:mhalo_mprog}
\end{figure*}

\subsubsection{Constraints on the dark matter density profile}\label{sec:density}
In this section, we present the streams' constraints on the radial density profile of the dark matter halo of m12m, following the same analysis as for the two streams evolved 
in analytic potentials presented in \citet{NibPear2025} and the mass enclosed profile. 
We have
fixed the baryonic component of m12m (see Figure \ref{fig:baryons-fitted}), 
and note that an incorrect baryonic fit in observations could bias the inferred dark matter radial density.

In Figure \ref{fig:radialprofile}, we show the dark matter density profiles of models sampled within the 68\% (dark) and 95\% (light) confidence regions of
the posteriors for the halo mass, $r_s$, $\gamma$, and $\beta$ for each stream. We also plot the median of the samples (colored lines). The black line shows the true dark matter density from our Zhao fit to
the FIRE m12m $z=0$ particles (Figure \ref{fig:zhao_z0}), and the gray region shows the prior range of our sampled parameter space in
log$_{10}M_{\rm halo}$, $r_s$, $\gamma$, and  $\beta$. 

All three streams constrain the radial density profile  better than the prior range at all radii. The constraints  widen toward the halo center, where the inner
density slope is only weakly constrained. 
For the S-shaped and Curlicue streams, the true profile lies within the 68\% region at all radii, and the Curlicue stream produces the tightest constraints on the outer slope, which is also reflected in the posteriors of the  $\beta$ parameter in the corner plots presented in Appendix \ref{sec:corner}. For the Massive stream, the median underestimates the
density at large radii, and the truth falls outside the 68\% region at large radii, though it remains within the 95\% region. This can also be seen from the preference for a higher $\beta$-value, and therefore a steeper outer slope, in the posteriors for the Massive stream run as compared to the true value (Figure \ref{fig:corner_inner}).
The median profiles at small radii of all three streams lie above the truth, reflecting a preference for cuspier profiles than the cored center of m12m.
We found one specific mode with our fit to the stars and dark matter from the FIRE particles described in Section \ref{sec:methods}, but there could be others, which could affect the inner slope. As our 68\% region includes the truth for the S-shaped and Curlicue streams, \texttt{X-Stream} has found a viable solution.

\begin{figure*}
    \centering
    \includegraphics[width=\textwidth]{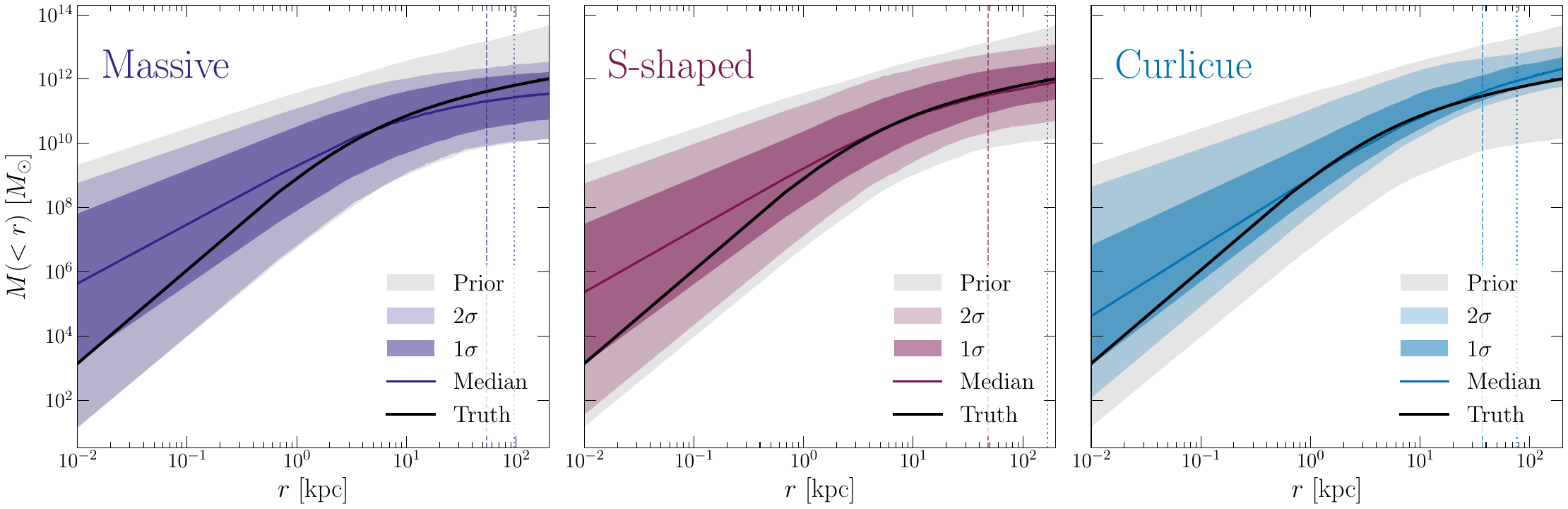}
    \caption{Constraints on the enclosed dark matter mass profiles of the host halo as a function of distance from its center for the massive (left), S-shaped (middle), and Curlicue (right) streams. The posterior sampling, colors, and lines are the same as in Figure \ref{fig:radialprofile}. The Curlicue stream shows the strongest constraints on both the inner and outer region which is also reflected in the posteriors in Figure \ref{fig:corner_curlicue}.  }
    \label{fig:menclosed}
\end{figure*}

\subsubsection{Halo and progenitor mass constraints}\label{sec:mass}
Several of our sampling conditions set a lower limit on the halo mass: for a fixed progenitor location, if the halo mass is too low, the tidal radius exceeds our stripping threshold and the progenitor does not strip (Section \ref{sec:xstream}), or the progenitor becomes unbound entirely. Our sampler also penalizes streams that are shorter than the control points. Thus, to produce a long enough stream if the integration time is too short, we need a more massive halo. This also sets a lower limit on the halo mass, since streams typically do not remain coherent and observable for much longer than 4 Gyr \citep{Mancillas2019}. 

The width  of the stream sets an upper limit on the halo mass. As discussed in \citet{NibPear2025}, for a fixed range of progenitor mass priors, a too massive halo can produce a stream that is too narrow compared to the control points. 
The halo mass is, however, degenerate with the progenitor mass and the
integration time. As discussed above, a shorter integration time requires a more massive halo for the stream to grow long enough to cover the data, and a more massive progenitor requires a shorter integration time, since its stars phase mix more rapidly. We bound these degeneracies with priors: the progenitor mass range can realistically be set from the observed surface brightnesses, and the length of the stream sets priors on the dynamical age (e.g., the observed tails are typically a few Gyr old). We additionally use physics informed velocity priors, excluding orbits that plunge to the center or escape radially, since neither forms a stream. These velocity priors further restrict the allowed halo masses at a given progenitor velocity.

In Figure \ref{fig:mhalo_mprog}, we show the posterior distributions of the progenitor and halo mass for each stream. The truth lines show the fitted Zhao halo scale mass to the FIRE m12m dark matter particles  (log$_{10}(M_{\rm halo}/{\rm M}_\odot) =$ 11.1, Table \ref{tab:BFEfit}) and the present-day progenitor stellar masses, including any stellar mass in the present-day streams (Table \ref{tab:streams}).

The three streams constrain the masses to varying degrees. 
For the Massive stream (left), the halo mass posterior is nearly flat across the prior range, and the progenitor mass posterior rises toward the upper prior boundary. The posterior contours are not closed, but there is a strong degeneracy between $M_{\rm halo}$ and $M_{\rm prog}$. This degenerate band includes the true halo and progenitor mass. This degeneracy arises from the tidal radius condition (Section \ref{sec:methods}): a more massive progenitor requires a stronger host tidal field to strip. 

For the S-shaped stream (middle), both posteriors are broad. The halo mass posterior peaks near the truth but remains nearly flat across most of the prior range, providing no closed constraint. The progenitor mass posterior sets a lower limit of log$_{10}(M_{\rm prog}/{\rm M}_\odot) >$ 7.9 at 68\% confidence, with the truth well within the allowed range. There is a strong degeneracy between $M_{\rm halo}$ and $M_{\rm prog}$. This degenerate band includes the true halo and progenitor mass.

For the Curlicue stream (right), the progenitor mass posterior is closed and peaks near the truth, with a median and 68\% interval of log$_{10}(M_{\rm prog}/{\rm M}_\odot) = $9.0$^{+0.44}_{-0.44}$ (truth: log$_{10}(M_{\rm prog}/{\rm M}_\odot) = 8.79$). The halo mass posterior peaks close to the truth, with log$_{10}(M_{\rm halo}/{\rm M}_\odot) = $11.3$^{+0.84}_{-0.72}$ (truth: 11.06), and declines toward higher masses: the 68\% confidence region is closed, while the 95\% region extends to the upper prior boundary at log$_{10}(M_{\rm halo}/{\rm M}_\odot) = 13$. 
The Curlicue stream is the only stream whose morphology bounds the halo mass from above. 

For each of the three streams, we also ran \texttt{X-Stream} where we fixed the radial velocity of the progenitor at present day to the true value from the simulation. 
In Section \ref{sec:orbit}, we discussed that for the fixed radial velocity run for the Massive stream, the velocity posteriors produced closed contours rather than lower limits, and the $y_{\rm prog}$ degeneracy breaks, giving a lower limit for the positive mode (progenitor in front of the host galaxy). The full corner plot for this run (Figure \ref{fig:corner_inner_vr}) places lower limits on halo mass and progenitor mass, whereas the run without a fixed radial velocity did not (Figure \ref{fig:mhalo_mprog}, left). Thus, fixing the radial velocity of the progenitor tightened the lower range of the joint constraint on the progenitor and halo mass, with the truth included in the degeneracy band.

For the S-shaped \texttt{X-Stream} run with fixed radial velocity (see Figure \ref{fig:corner_outer_vr}), where the $y_{\rm prog}$ degeneracy was not broken and where there were still only lower limits on the velocity parameters, there is no significant difference in the joint posterior distribution of the progenitor and halo mass parameters, which still show a strong degeneracy across the prior range, with the truth included in the degenerate band.

From the Curlicue \texttt{X-Stream} run with fixed radial velocity (see Figure \ref{fig:corner_curlicue_vr}), where the degeneracy in $y_{\rm prog}$ was broken, the progenitor and halo mass posteriors are again closed and peak near the truth. The joint posterior of the progenitor and halo mass parameters closely resembles that of the run without a fixed radial velocity. 
See Section \ref{sec:futureobs}, for a discussion of how these results compare to the findings in \citet{pearson2022b}, where they used a fixed progenitor mass to place limits on  Centaurus A's mass from a single stream, and see Appendix \ref{sec:corner} for a discussion of all parameters in the full corner plots.

In Figure \ref{fig:menclosed}, we show  constraints on the enclosed dark matter mass profile, $M(<r)$. All three streams constrain the enclosed mass better than the prior at large radii where the streams predominately reside, though there is little information in enclosed mass at small radii. For all three streams, the true profile lies within the 68\% region. The Curlicue stream shows the strongest constraints, with a 68\% interval spanning approximately 0.6 dex at $r =$ 200 kpc, compared to $\sim 1.5$ dex and $\sim1.2$ dex for the Massive and S-shaped stream, respectively. 
For the Massive stream, the median falls below the truth at large radii, consistent with its underestimated density at those radii in Figure \ref{fig:radialprofile}. 
Note that since the enclosed mass is an integrated quantity, errors accumulate from all radii $<r$. The $M(<r)$ constraint therefore is less constrained than the radial profile, despite sampling over the same parameters.  Additionally, the Massive and S-shaped streams contain no information on the inner mass profile, since they are not radially plunging and have a limited spatial extent compared to the Curlicue stream.

\section{Discussion}\label{sec:discuss}
In this section, we discuss differences between each of the three streams and how including more extended debris affects our inference (Section \ref{sec:extendeddebrisdiscus}). We place our findings in context of the future of extragalactic stream science (Section \ref{sec:futureobs}), and discuss the limitations of our approach (Section \ref{sec:limitations}). 

\subsection{Fitting the diffuse and longer parts of the streams}\label{sec:extendeddebrisdiscus}
The curlicue stream is the most constraining across all parameters. This is likely due to its track curvature, length, and phase coverage (see \citealt{nibauer2023,Chemaly2026,Chemaly2026b,Wu2026,starkman2026} and Eq. 4 in \citealt{NibPear2025}). 
The stream fits sampled from
within  68\% of the posterior distributions for the Curlicue stream,  visualized in Figure \ref{fig:truevsfit}, correctly included the extended wrap of the stream seen in FIRE (upper right panel in Figure \ref{fig:true6D}), despite the red control points not including this under-dense region of the stream. The fact that \texttt{X-Stream} was able to find the correct mode is likely due to an interplay between the position of the progenitor and the gap between the control points at $x,z\sim 25,25$ kpc. 
For the other two streams, only the red control points, tracing the densest parts of the streams, were reproduced in the fits visualized in Figure \ref{fig:truevsfit}. 

To test how including the more diffuse, longer, extended parts of the streams affects our inference, we repeat the  \texttt{X-Stream} runs with control points extending further along the leading and trailing arms of the FIRE streams. In Section \ref{sec:Results}, we selected 10\% of the densest points with SB $>29.5$ $\rm{mag} \, \rm{arcsec}^{-2}$ (see Figure \ref{fig:projections}). Here we instead use 30\% of the densest points. Note that for the Curlicue stream,  this means that the stream looks like a fully filled oval, without the gap at $x,z\sim 25,25$ kpc. This configuration  is not constraining \citep[see e.g.,][]{Wu2026}, and therefore we focus on the S-shaped and Massive streams in this section. 

\begin{figure*}
    \centering
    \includegraphics[width=\textwidth]{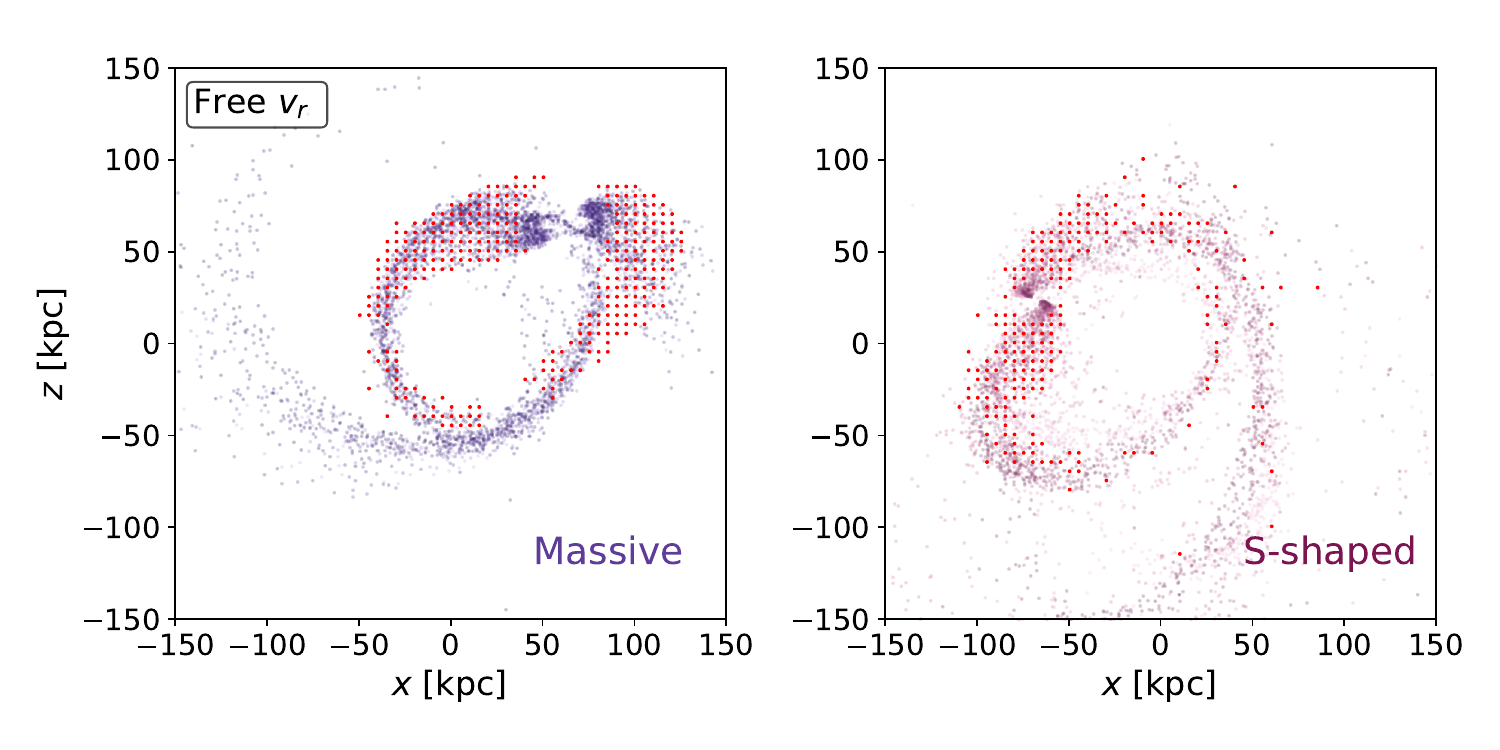}
    \caption{Same as top panel of Figure \ref{fig:truevsfit}, but for \texttt{X-Stream} runs with more extended control points for the Massive (left) and S-shaped (right) streams. We do not show the Curlicue runs,
    as the more extended control points caused the stream to form a single ring-like structure, which we could not constrain.
    In each panel we overplot five streams randomly selected within the 68\% posterior confidence regions presented in  Appendix \ref{sec:extendedpoints}. 
     }
    \label{fig:extendedpoints}
\end{figure*}

In Figure \ref{fig:extendedpoints}, we show the red control points from the 30\% cut used as input in the new \texttt{X-Stream} runs. %
In observations we would not be able to disentangle each stream from the other (see Figure \ref{fig:sim}).  
The runs presented here are instead meant to test how our inference is affected if we include more extended parts of streams. 

In Figure \ref{fig:extendedpoints} we again show five streams  randomly  selected  from within the 68\% posterior confidence regions for the new \texttt{X-Stream} runs for the Massive (left) and S-shaped (right) streams.  As expected from the new input data, the randomly  selected streams are longer, covering the extended control points. 
We present the full corner plots in Appendix \ref{sec:extendedpoints}, and briefly summarize the main findings below.

For the Massive stream,  the posterior distribution for $y_{\rm prog}$ is flat, and $\eta_r$ is now a closed contour with a bimodial distribution instead of a lower limit. The truth values for the $\eta_r$ and $\psi$ parameters are close to the lower edge of the posterior values as they were before. While there are still no strong limits on  $M_{\rm halo}$ and $M_{\rm prog}$, 
we again see a strong degeneracy between $M_{\rm halo}$ and $M_{\rm prog}$. This degenerate band includes the true halo and progenitor mass. 
As discussed in Section \ref{sec:orbit}, the fact that the true velocity values are still at the lower edge of the posteriors can be due to the red control points having lower curvature than the true 6D stream, when modeled without time-dependent effects (see Figure \ref{fig:true6D}). 

For the S-shaped stream, the orbital constraints are now closed contours, as opposed to lower limits, with the true values recovered within the 68\% region for the S-shaped stream. There is a lower limit on the halo mass, since there are less ways to get the correct morphology of the red points, once we include the wrap with specific curvature (see Eq. 4 in \citealt{NibPear2025}). 
We do not see an upper limit on halo mass. 
For the S-shaped stream, including more extended debris  yields tighter constraints.

\subsection{Implications for Extragalactic Stream Modeling}\label{sec:futureobs}
Fitting the three m12m streams with \texttt{X-Stream}, assuming a static potential, works well even in a cosmological setting: this includes both the densest, most recently stripped parts of the Curlicue streams and the more extended parts of the S-shaped stream. Correctly capturing episodic stripping, where dwarfs strip close to pericenter, is essential to this modeling. In future observations, selecting isolated progenitors, lacking visible companion satellites, other streams, or signatures of recent mergers, reduces the risk of time-dependent perturbations like those affecting the Massive stream. 
Disentangling overlapping streams within a single halo is difficult in observations \citep{bell2026}, which would also be the case for the three m12m streams (Figure \ref{fig:sim}). 
\texttt{X-Stream}  has already been applied to an  extragalactic globular cluster stream \citep{Holm2026}, and {\it Euclid} \citep{racca2016} and  {\it Roman} \citep{pearson2019, pearson2022a} might detect many more such streams.  

An ideal target also shows the progenitor itself alongside the recently stripped, bright part of the stream, but progenitors are rarely visible \citep{sola2025}. When a progenitor location is unknown, we can marginalize over it, though this is expensive. A cheaper path is to first apply curvature-based methods \citep{nibauer2023,Wu2026,starkman2026} to select candidates and inform priors, then run generative modeling. 
Note that in this paper, the inference on the Curlicue stream worked well despite the progenitor being fully disrupted at $z=0$ and located from the median position and velocity of the 500 particles closest to the highest-density point in the unwrapped stream. 

Radial velocities along the stream, from globular clusters \citep[e.g.,][]{mueller2025},  planetary nebulae \citep{Valenzuela2026}, or multi-object spectroscopy \citep{Toloba2016}, can help the sampler find the correct solution, break degeneracies, and constrain the allowed orbital space \citep{pearson2022b}. Radial velocities are natural targets for follow-up, but our runs with fixed radial velocity show that the information gained from a single radial velocity depends on the individual stream (see Section \ref{sec:Results}). \citet{pearson2022b} found that one radial velocity measurement of the Dwarf 3 stream orbiting Centaurus A placed a lower limit on the host halo mass, using a fixed progenitor mass rather than the wider prior range adopted here. Because they had no halo mass versus progenitor mass degeneracy, that radial velocity measurement constrained halo mass directly. In this paper, we also model the stream width, whereas \citet{pearson2022b} fit only the stream track. Even with width included, the constraining power of a radial velocity measurement is stream dependent.

Deeper data and detailed sky subtraction of streams on certain orbits reveal previous wraps of tidal debris \citep{sola2025, bell2026, Ogami2026}. Our analysis of extended debris shows that these previous wraps can help constrain the host halo's present-day parameters, even without adding time-dependent effects to the modeling. This works well for the S-shaped stream, which shows no evidence of a recent encounter with another satellite or of significant orbital decay within the last 6 Gyr (Figure \ref{fig:infall}).

Alternatively, a population-level analysis of many streams, adopting an average growth rate for hosts of a given mass, could constrain halo properties statistically. Curvature-based methods are better suited to this population-level work \citep[see e.g.,][]{nibauer2023,Wu2026,starkman2026} than our per-system sampler, although posteriors from many systems could also be combined through generative forward modeling \citep[see e.g.,][]{Chemaly2026b}.

\subsection{Limitations}\label{sec:limitations}
In \citet{NibPear2025}, we discussed limitations of \texttt{X-Stream} itself, including the particle-spray method, the runtime, the unknown progenitor location, the assumption of a spherical halo, the assumption of knowing the distance to the host without errors, and the degeneracies inherent to fitting extragalactic streams from imaging alone, which also appear in the corner plots in this work (Appendix \ref{sec:corner}). Here, we focus on limitations specific to
applying our method to a cosmologically evolved system.

Section \ref{sec:nofit} documents two time-dependent effects likely contributing to the Massive stream's offset: a visually apparent encounter perturbing its leading arm at $t\sim -0.8$ Gyr, and  dynamical friction on the progenitor's orbit (Figure \ref{fig:infall}). 
Neither is included in our static model. 
Throughout the paper we also make the assumption of Lagrange point stripping in the particle-spray simulations, which is inappropriate for lower mass-ratio mergers. Whether this affects the Massive stream could be tested with N-body simulations of the infall, which we leave to future work.

The simulation we have utilized, m12m, is one example of a halo, which has not had  a major merger \citep{Panithanpaisal2021,Horta2023}, nor any significant change to its halo shape over that past 5 Gyr \citep{arora2025}. A wider study across halos and across cosmological simulations, such as other FIRE and Auriga halos \citep{riley2025,shipp2025}, can test how representative our results are. FIRE streams also differ systematically from observed Milky Way streams, where simulated streams have larger pericenters and apocenters than observed Milky Way streams \citep{Li2022, Shipp2023}, reflecting over-disruption in simulations or progenitors puffed up by stellar feedback in FIRE. Thus the streams analyzed in this paper do not reflect all streams we will detect in upcoming data sets. 

Our condition that stars strip only where $r_J < 3 \times r_e$ requires choosing a threshold. We chose a factor of 3 because it confines stripping to pericenter passages. A different threshold, or full N-body runs could change our choice of the stripping threshold. 

In this work, we did not include the gas to the fixed potential when modeling the baryons. While this might affect the details of the radial profile and mass constraints at small radii, it should not affect the overall trends in constraints for each stream. 

With m12m, we can check our inference against the true simulation values. In observations we cannot, and have limited ways of knowing ahead of time whether a given stream is well described by our model assumptions, or biased like the Massive stream, without combining with other constraints. We have shown that we can recover the true parameters under the assumption of a spherical halo. Streams are sensitive to halo flattening \citep{nibauer2023,Wu2026,starkman2026,Chemaly2026b}, and the m12m halo is in fact flattened in dark matter \citep[see solid blue line in][Fig. 8, right]{Vargya2022}, which would lead to a bias towards higher enclosed halo masses at a given radius \citep[see e.g., Fig. 12.12 in][]{Bovy2026}.
In future work, we plan to relax our spherical assumption with more flexible potential modeling.

\section{Conclusion}\label{sec:conclusion}
We applied the \citet{NibPear2025} \texttt{X-Stream} sampler to three accreted dwarf galaxy streams in the FIRE m12m cosmological zoom-in simulation, treated as an  extragalactic system. We tested how well we recover the true present-day halo and progenitor parameters under the assumption of a static halo with and without radial velocity information. Our main conclusions are summarized below:

\begin{itemize}

\item The densest debris from the three m12m streams would be observable in wide field {\it Euclid} data, but the extended diffuse debris would be difficult to disentangle between the three streams (Figure \ref{fig:sim}).

\item A tidal radius condition is crucial when modeling cosmologically evolved dwarf galaxy streams with particle-spray techniques. Dwarfs strip near pericenter (e.g., \citealt{bonaca2025}), and if stars are modeled to strip uniformly along the orbit this can produce incorrect stream morphologies. Starting from the progenitor's true 6D phase-space coordinates, our static potential model reproduces the most recently stripped debris for two of the three FIRE m12m streams, but only when the stripping time depends on the tidal radius (Figure~\ref{fig:true6D}).

 \item  Even under the assumption of a static halo, \texttt{X-Stream} produces closed constraints and recovers the true orbital parameters for the Curlicue stream using only the morphology of the observable parts of the stream. For the S-shaped stream, when including more extended debris, \texttt{X-Stream} also produces closed constraints and recovers the true orbital parameters, while yielding  upper limits on velocity if only the densest part of the stream is used as input data.  
  \texttt{X-Stream} produces biased orbital limits for the Massive stream, consistent with the unmodeled time-dependent effects discussed in Section \ref{sec:limitations} or from the fact that this is a lower mass-ratio merger.

\item The S-shaped and Curlicue streams recover the radial density profile of m12m within the 68\% confidence regions at all radii, with the Curlicue stream placing the tightest constraints on the outer density slope, $\beta$.  The Massive stream posteriors deviate at the 95\% level towards steeper outer radial density. 

\item If we include only the densest parts of each stream, only the Curlicue stream places strong constraints on the halo mass and progenitor mass, while the other two streams only produce weak constraints on the enclosed halo mass. When including more extended and diffuse parts of the streams as input to \texttt{X-Stream}, we can also place a lower limit on the halo mass from the S-shaped stream, whereas the Massive stream's halo and progenitor mass remain unconstrained even with the extended debris included. 

\item 
\citet{pearson2022b} found that for a fixed progenitor mass, one radial velocity  constrained the halo mass for the DW3 stream in Centaurus A. However, in our work, where we have a range of allowed progenitor masses, we find that fixing the radial velocity of the progenitor does not break degeneracies or tighten halo mass constraints uniformly across streams. Fixing the radial velocity breaks the $y_{\rm prog}$ degeneracy for the Curlicue and Massive streams, but not for the S-shaped stream. Fixing the radial velocity also tightens the joint posterior on halo mass and progenitor mass only for the Massive stream, while the S-shaped and Curlicue stream posteriors show no such tightening.

\end{itemize}

Upcoming surveys, including {\it Euclid} \citep{racca2016,starkman2026}, Rubin \citep{ivezic2019}, {\it Roman} \citep{spergel2015}, {\it ARRAKIHS} \citep{guzman2022}, and LIGHTS \citep{Zaritsky2024,Zaritsky2026A}, will deliver deep imaging of extragalactic stellar streams. Our results show that imaging of recently stripped debris from dwarf progenitors can constrain present-day dark matter halo properties despite the halo's time-dependence. A single constraining stream, such as Curlicue, can serve as a prior on halo mass for other streams in the same system. In real observations, the main challenge will be identifying which streams are least disturbed by time-dependent effects. Streams from lower-mass progenitors limit dynamical friction, and systems without recent mergers or satellites are a good starting point.

\begin{acknowledgments} 
This work was supported by a research grant (VIL53081) from VILLUM FONDEN. This work was also co-funded by the European Union (ERC, BeyondSTREAMS, 101115754) grant. Views and opinions expressed are however those of the author(s) only and do not necessarily reflect those of the European Union or the European Research Council. Neither the European Union nor the granting authority can be held responsible for them. AA acknowledges support from Gordon and Betty Moore foundation. 
The authors used Claude (Anthropic) as sparring partners to refine plots and debug, and to improve clarity and conciseness during manuscript revision in the editing stage. The authors take full responsibility for the content of this work.
The Tycho supercomputer hosted at the SCIENCE HPC center at the University of Copenhagen was used for supporting this work. 
\end{acknowledgments}

\software{astropy \citep{2013A&A...558A..33A,2018AJ....156..123A}, streamsculptor \citep{Nibauer2025a}, nautilus \citep{nautilus}, JAX \citep{Jax2018}, gala \citep{gala}, py-ananke \citep{Thob2024}.}

\appendix
\section{Fit to star particles and true orbits in FIRE m12m}\label{sec:baryons}
\begin{figure}
    \begin{minipage}{0.5\columnwidth}
    \centering
    \includegraphics[width=\linewidth]{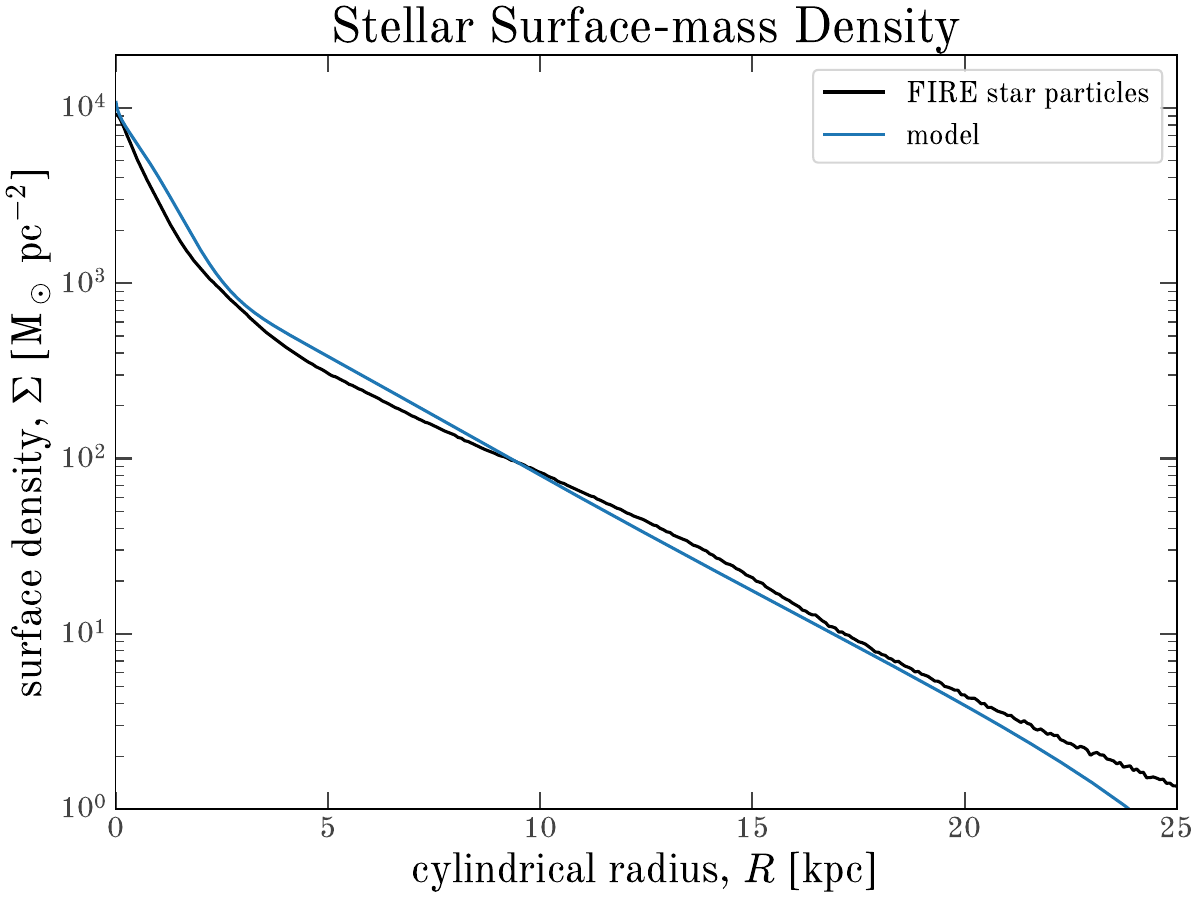}
    \caption{The surface-mass density profile $\Sigma(R)$ of star particles from the m12m simulation at redshift $z=0$ (black) and the same computed from our best-fit, analytic potential model representation of the stars (blue line). Our full procedure for fitting the star particle mass distribution is described in Section~\ref{sec:potential}.}
    \label{fig:baryons-fitted}
    \end{minipage}
\end{figure}

Throughout the paper, we fixed the potential of the stars in FIRE m12m at $z=0$. We describe the procedure for obtaining this fit in Section \ref{sec:potential}. 
In Figure \ref{fig:baryons-fitted}, we present our best fit to the star particles at  $z=0$ (blue) and compare to the surface-mass density profile from the star particles in the m12m simulation at  $z=0$ (black).  

\begin{figure}
    \begin{minipage}{0.5\columnwidth}
        \centering
        \includegraphics[width=\linewidth]{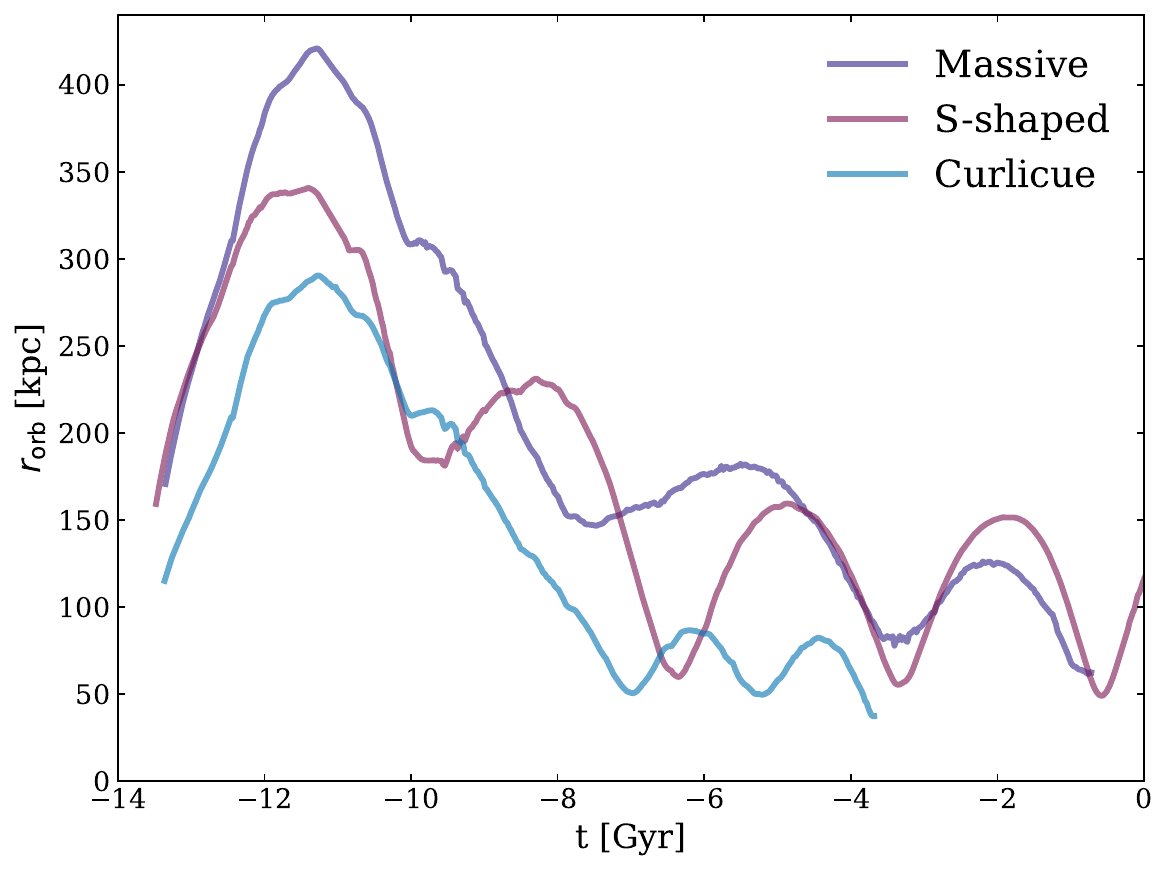}
        \caption{The galactocentric radius (orbit) evolution of each progenitor evolved in the FIRE m12m cosmological simulation for the massive (purple), S-shaped (pink), and Curlicue (blue) stream progenitors. $t=0$ is present day at $z=0$. }
        \label{fig:infall}
    \end{minipage}
\end{figure}

Figure \ref{fig:infall} shows the galactocentric orbital evolution of each progenitor from 
$t=-13.8$ Gyr to present day. 
The S-shaped progenitor remains intact today, the Massive stream progenitor fully disrupts at $t\sim -0.8$ Gyr, and the Curlicue progenitor disrupts at $t\sim -3.8$ Gyr. 
The orbits evolve due to a combination of growth of the host, mass loss of the satellites, interaction amongst the satellites, and dynamical friction \citep{Santistevan2023}. 
At present day, the Massive stream contains 1.8 times the stellar mass of the Curlicue stream, and 3.2 times that of the S-shaped stream and its progenitor. 
The S-shaped and Curlicue progenitors also show early orbital decay but stabilize within the last 6 Gyr. Once the Curlicue progenitor fully disrupts, its stream no longer experiences dynamical friction.

\section{Summary of corner plots for all \texttt{X-Stream} runs}\label{sec:corner}

Here we present and briefly summarize the full corner plots for the three m12m streams for runs with 10 free parameters, and runs with 9 free parameters, where the radial velocity of the progenitor was fixed. 

\begin{figure*}
    \centering
    \includegraphics[width=\textwidth]{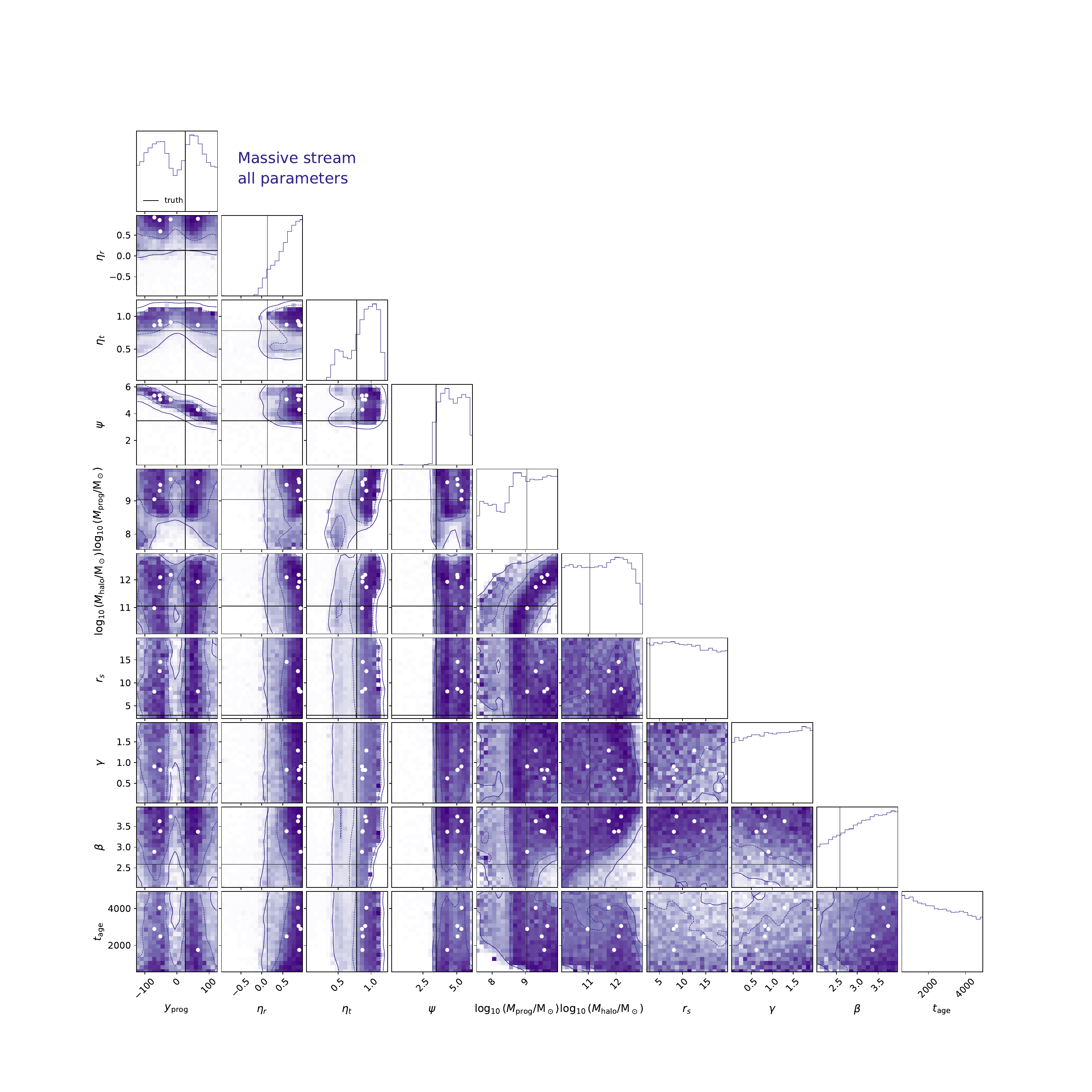}
    \caption{Constraints on the 10 free parameters for the Massive stream. 
    Purple dashed contours show the 68\% levels, the solid purple lines show the 95\% levels in both the contours and 1D histograms. The black lines show the true parameters. The white points show  five randomly selected good fits within the 68\% credible region. We  visualize the streams corresponding to these points in Figure \ref{fig:truevsfit}.
    }
    \label{fig:corner_inner}
\end{figure*}

Figure \ref{fig:corner_inner} shows the constraints for the Massive stream \texttt{X-Stream} run with 10 free parameters. Contours show the 68\% and 95\% confidence regions, black lines show the true parameters, and white points show solutions within the 68\% region.
The $y_{\rm prog}$ posterior is bimodal, with the true value ($y_{\rm prog}=26$ kpc) falling at the lower edge of the positive mode. The bimodality reflects the degeneracy between a progenitor located in front of or behind the host galaxy. The $\eta_r$ posterior runs into the upper prior boundary, giving a limit on $\eta_r$. The true value,  falls outside the 68\% region but within the 95\% region. $\eta_t$ is well recovered, with the truth  inside the 68\% region. The true $\psi$ value sits at the lower edge of its posterior.

The halo mass posterior is nearly flat across the prior range, and the progenitor mass posterior rises toward the upper prior boundary.
This is consistent with the biased orbital parameters discussed above: an unconstrained velocity leaves the mass required to keep the stream bound essentially unconstrained as well. As discussed in Section \ref{sec:mass}, there is a strong degeneracy between $M_{\rm halo}$ and $M_{\rm prog}$. This degenerate band includes the true halo and progenitor mass, and arises because a more massive progenitor requires a stronger host tidal field to strip. \citet{NibPear2025} did not include a tidal radius condition, and therefore did not find this degeneracy.

The parameter $\beta$ pushes toward the upper prior boundary, consistent with the general degeneracy between a steeper outer slope and higher halo mass, and yields only a 68\% lower limit. $r_s$, $\gamma$, and $t_{\rm age}$ show weak to no constraints across the prior range.
As discussed in \citet{NibPear2025}, the degeneracy between $\beta$ and  halo mass, where a steeper outer radial profile (higher $\beta$) requires a higher halo mass, arises since a higher $\beta$ truncates the intermediate-to-outer density, requiring a higher overall mass amplitude to match the input data.

\begin{figure*}
    \centering
    \includegraphics[width=\textwidth]{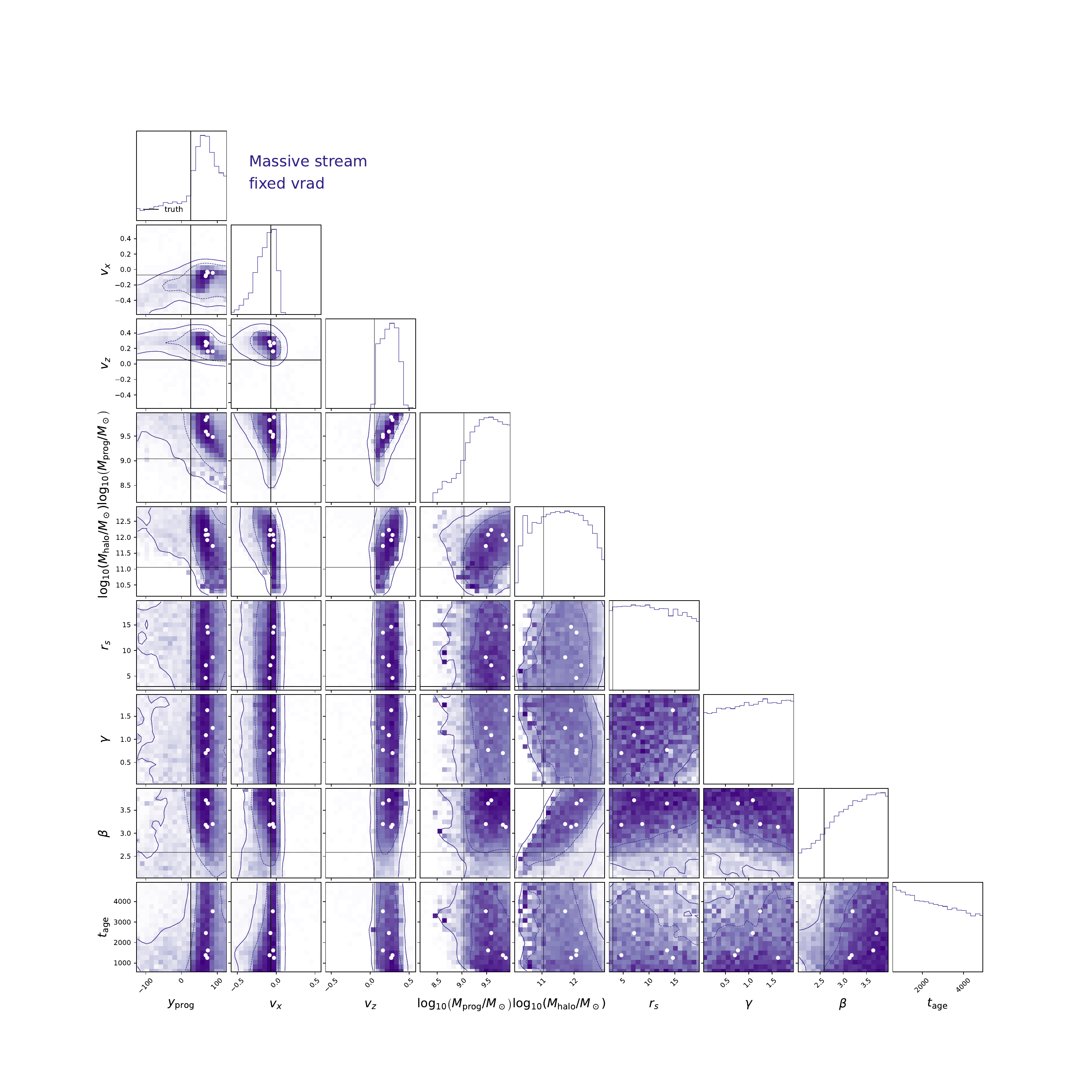}
    \caption{Constraints on the 9 free parameters for the Massive stream in the \texttt{X-Stream} run, where we fix the radial velocity ($v_y$) of the progenitor to the true value from FIRE. Note that the velocity sampling is now done in $v_x$ and $v_z$. 
    The lines and points are the same as in Figure \ref{fig:corner_inner}.
    }
    \label{fig:corner_inner_vr}
\end{figure*}

Figure \ref{fig:corner_inner_vr} shows the constraints for the Massive stream \texttt{X-Stream} run with 9 free parameters, where the radial velocity of the progenitor was fixed to the true value. Note that velocity sampling is now done in $v_x$ and $v_z$, so these are not directly comparable to $\eta_r$, $\eta_t$, and $\psi$ in the free-velocity run.

Fixing the radial velocity breaks the $y_{\rm prog}$ bimodality seen in Figure \ref{fig:corner_inner}, though the true value remains offset from the posterior, sitting at the lower edge of the 68\% contour. The $v_x$ and $v_z$ posteriors are now closed, with the true $v_x$ within the 68\% region and the true $v_z$ at the lower edge of the 95\% region.

This closing of the velocity posteriors is accompanied by a corresponding tightening of the mass constraints: where $M_{\rm halo}$ and $M_{\rm prog}$ were unconstrained in the free-velocity run, fixing $v_{\rm rad}$ now yields a lower limit on $M_{\rm prog}$ (truth within the 95\% region) and a broad, closed 68\% constraint on $M_{\rm halo}$ that contains the truth. $\beta$ still prefers high values, with the truth outside the 68\% region but within the 95\% region. $r_s$, $\gamma$, and $t_{\rm age}$ remain weakly constrained in both runs.

\begin{figure*}
    \centering
    \includegraphics[width=\textwidth]{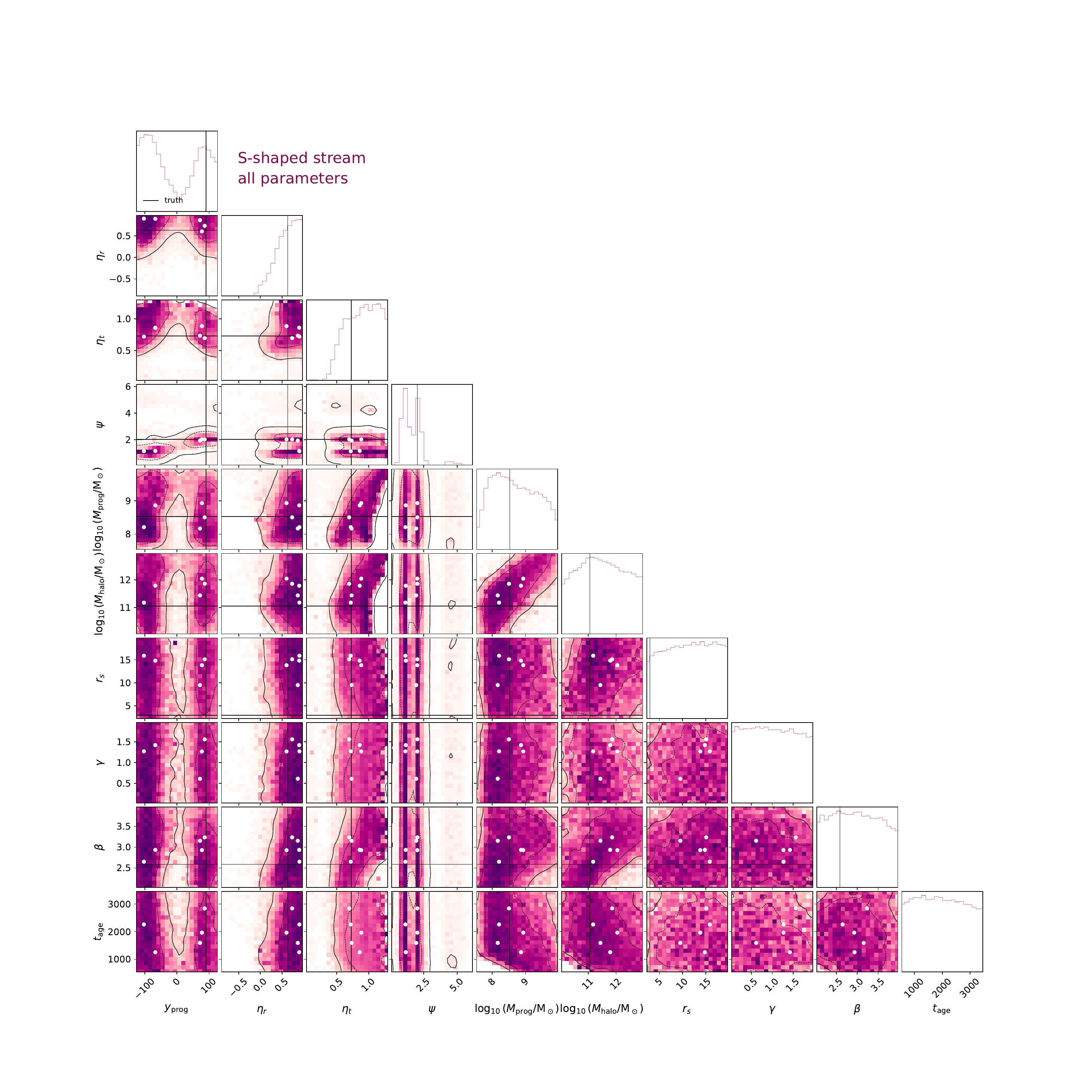}
    \caption{
    Constraints on the 10 free parameters for the S-shaped stream. 
    The lines and points are the same as in Figure \ref{fig:corner_inner}. 
    }
    \label{fig:corner_outer}
\end{figure*}

Figure \ref{fig:corner_outer} shows the constraints for the S-shaped stream \texttt{X-Stream} with 10 free parameters. We recover all 10 free parameters within the 68\% region, but the stream only weakly constrains the scale radius $r_s$, $\gamma$, $\beta$, and $t_{\rm age}$. $y_{\rm prog}$ shows a bimodality, where equally good fits exist for a progenitor located in front of or behind the host galaxy. Both $\eta_r$ and $\eta_t$ have lower limits, and $\psi$ is strongly constrained. There is a degeneracy between $\eta_r$ and the halo and progenitor masses, where faster moving streams (higher $\eta_r$) prefer higher halo and progenitor masses: a faster stream needs a higher halo mass to remain bound, and a higher halo mass in turn requires a higher progenitor mass to match the width of the input data. 
As for the massive stream, there is a strong degeneracy between $M_{\rm halo}$ and $M_{\rm prog}$, and between $\beta$ and $M_{\rm halo}$. 

\begin{figure*}
    \centering
    \includegraphics[width=\textwidth]{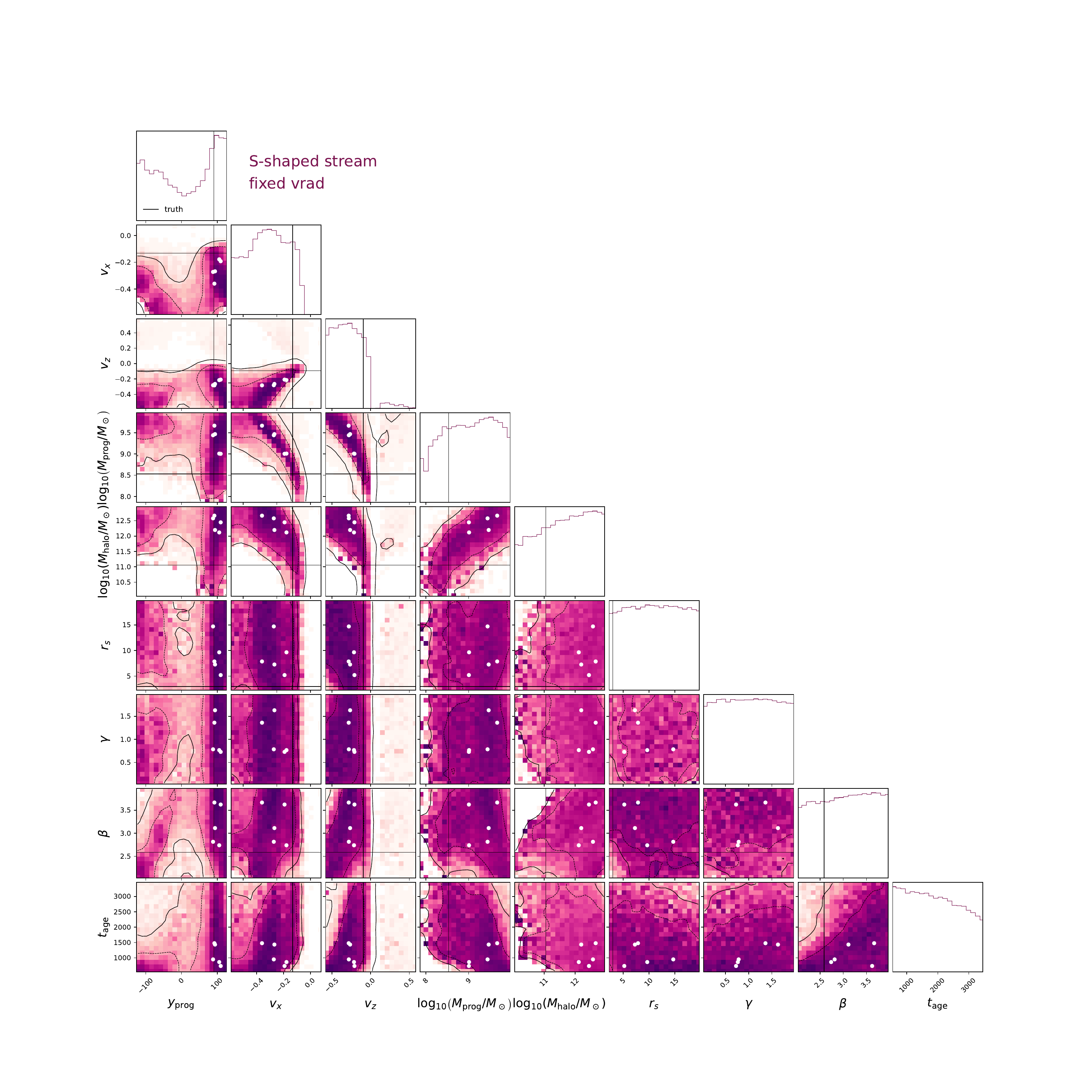}
    \caption{Constraints on the 9 free parameters for the S-shaped stream in the \texttt{X-Stream} run, where we fix the radial velocity of the progenitor ($v_y$) to the true value from FIRE. Note that the velocity sampling is now done in $v_x$ and $v_z$. The lines and points are the same as in Figure \ref{fig:corner_inner}.  
    }
    \label{fig:corner_outer_vr}
\end{figure*}

Figure \ref{fig:corner_outer_vr} also shows constraints for the S-shaped stream, now for an \texttt{X-Stream} run with the progenitor's radial velocity fixed and only 9 free parameters. As discussed in Section \ref{sec:orbit}, the degeneracy in $y_{\rm prog}$ persists despite fixing the radial velocity. We again recover all 9 free parameters within the 68\% region. $r_s$ and $\gamma$ remain only weakly constrained. The joint posteriors between $t_{\rm age}$ and $M_{\rm halo}$, $M_{\rm prog}$, and $\beta$ show that modes with the progenitor located behind the host galaxy (negative $y_{\rm prog}$) can only reproduce the input data with a high halo mass, progenitor mass, and velocity. Placing the progenitor farther behind the host requires a shallower outer slope. The degeneracy between $M_{\rm halo}$ and $M_{\rm prog}$ remains strong, and fixing the radial velocity does not tighten their joint posterior.

\begin{figure*}
    \centering
    \includegraphics[width=\textwidth]{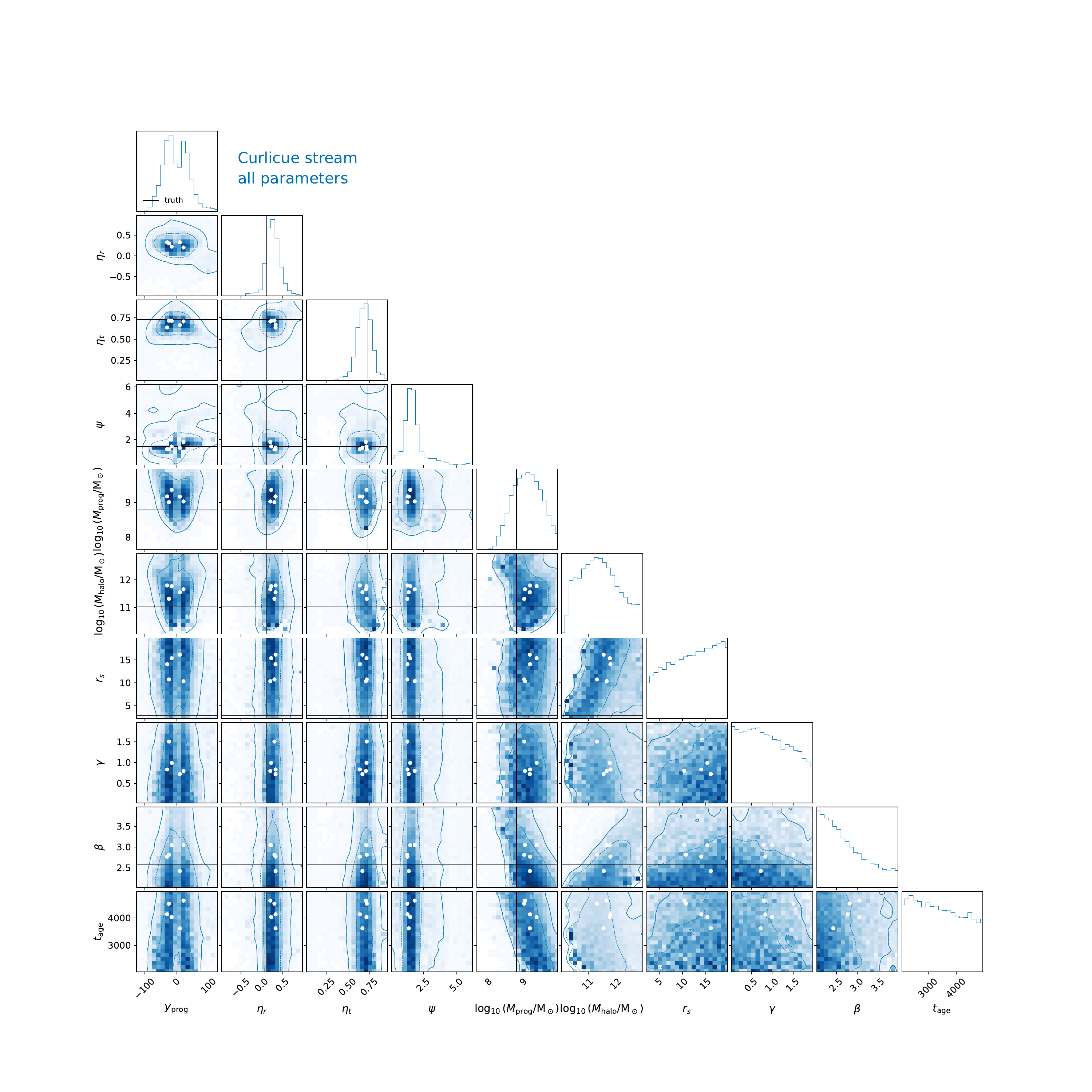}
    \caption{ Constraints on the 10 free parameters for the Curlicue stream. 
    The lines and points are the same as in Figure \ref{fig:corner_inner}.     
    }
    \label{fig:corner_curlicue}
\end{figure*}

Figure \ref{fig:corner_curlicue} shows the constraints for the Curlicue stream \texttt{X-Stream} run with 10 free parameters. 
Unlike the Massive and S-shaped streams, the joint posteriors of $\eta_r$, $\eta_t$, and $\psi$ are all closed and cluster tightly around the truth. The $y_{\rm prog}$ posterior remains bimodal, as for the other streams, reflecting the front-behind degeneracy in line-of-sight position.

The joint 68\% posterior of $M_{\rm halo}$ and $M_{\rm prog}$ is a closed contour rather than a  degenerate band: unlike the S-shaped and Massive stream, where the two masses trade off along an extended ridge with no upper bound, the Curlicue joint posterior is bounded in both dimensions and contains the truth. This is the only stream for which the mass constraint is closed in this way, consistent with the stream setting both a lower and an upper limit on the halo mass at the 68\% level.  As for the other two streams, we see a degeneracy between $\beta$ and $M_{\rm halo}$. The Curlicue stream is the only stream for which we see a strong degeneracy between $r_s$ and $M_{\rm halo}$, with larger halo masses requiring higher scale radii. There is also a degeneracy between  $t_{\rm age}$ and $M_{\rm prog}$: shorter integration times pair with higher progenitor and halo masses. \citet{NibPear2025} discuss this same degeneracy, showing that a lower-mass progenitor needs a longer integration time to produce a stream of the same length, since its stars escape at lower velocities. 

\begin{figure*}
    \centering
    \includegraphics[width=\textwidth]{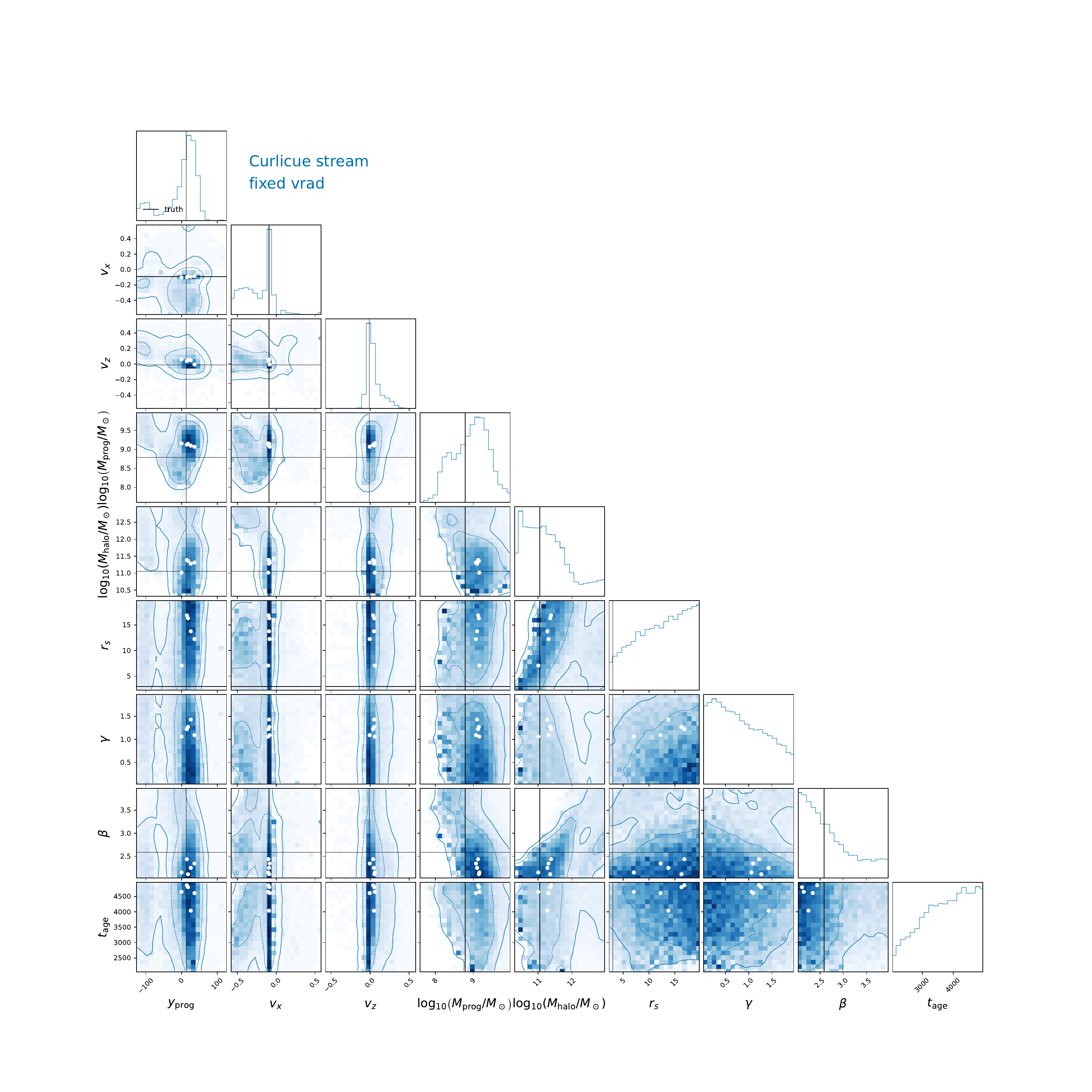}
    \caption{Constraints on the 9 free parameters for the Curlicue stream in the \texttt{X-Stream} run, where we fix the radial velocity of the progenitor to the true value from FIRE. Note that the velocity sampling is now done in $v_x$ and $v_z$. The lines and points are the same as in Figure \ref{fig:corner_inner}.
   }
    \label{fig:corner_curlicue_vr}
\end{figure*}

Figure \ref{fig:corner_curlicue_vr} shows the constraints for the Curlicue stream \texttt{X-Stream} run with 9 free parameters, where the radial velocity of the progenitor was fixed to the true value. As for the Massive stream, fixing the radial velocity collapses the $y_{\rm prog}$ bimodality into a single mode consistent with the truth. The $v_x$-$v_z$ joint posterior is sharply peaked, where the truth is included within the 68\% region. 

The $M_{\rm halo}$-$M_{\rm prog}$ joint posterior remains a closed contour of similar shape and extent to the free-velocity run described above. Fixing the radial velocity does not constrain it further, but only shifts it slightly. This is in contrast to the Massive stream, where fixing the radial velocity converts the mass posteriors from open, unconstrained bands into lower limits. The joint posteriors for the $r_s$-$M_{\rm halo}$ and $\beta$-$M_{\rm halo}$ are similar to the run with a free radial velocity. 

Overall, the Curlicue stream is unique among the three in producing a closed, bounded joint posterior on halo and progenitor mass from the morphology alone, and this joint constraint is set by the stream's shape rather than by the radial velocity: fixing the radial velocity tightens the orbital joint posteriors without tightening the mass joint posterior.

\section{Summary of \texttt{X-Stream} runs including extended debris}\label{sec:extendedpoints}
In this Section we present corner plots from our runs with \texttt{X-Stream} with 10 free parameters, where we use more extended control points as input data, discussed in Section \ref{sec:extendeddebrisdiscus} and shown in Figure \ref{fig:extendedpoints}. We use the same prior ranges as presented in Table \ref{tab:freeparams}, but now allow for up to 5 Gyr integration time for both streams, instead of a shorter integration time for the S-shaped stream.

\begin{figure*}
    \centering
    \includegraphics[width=\textwidth]{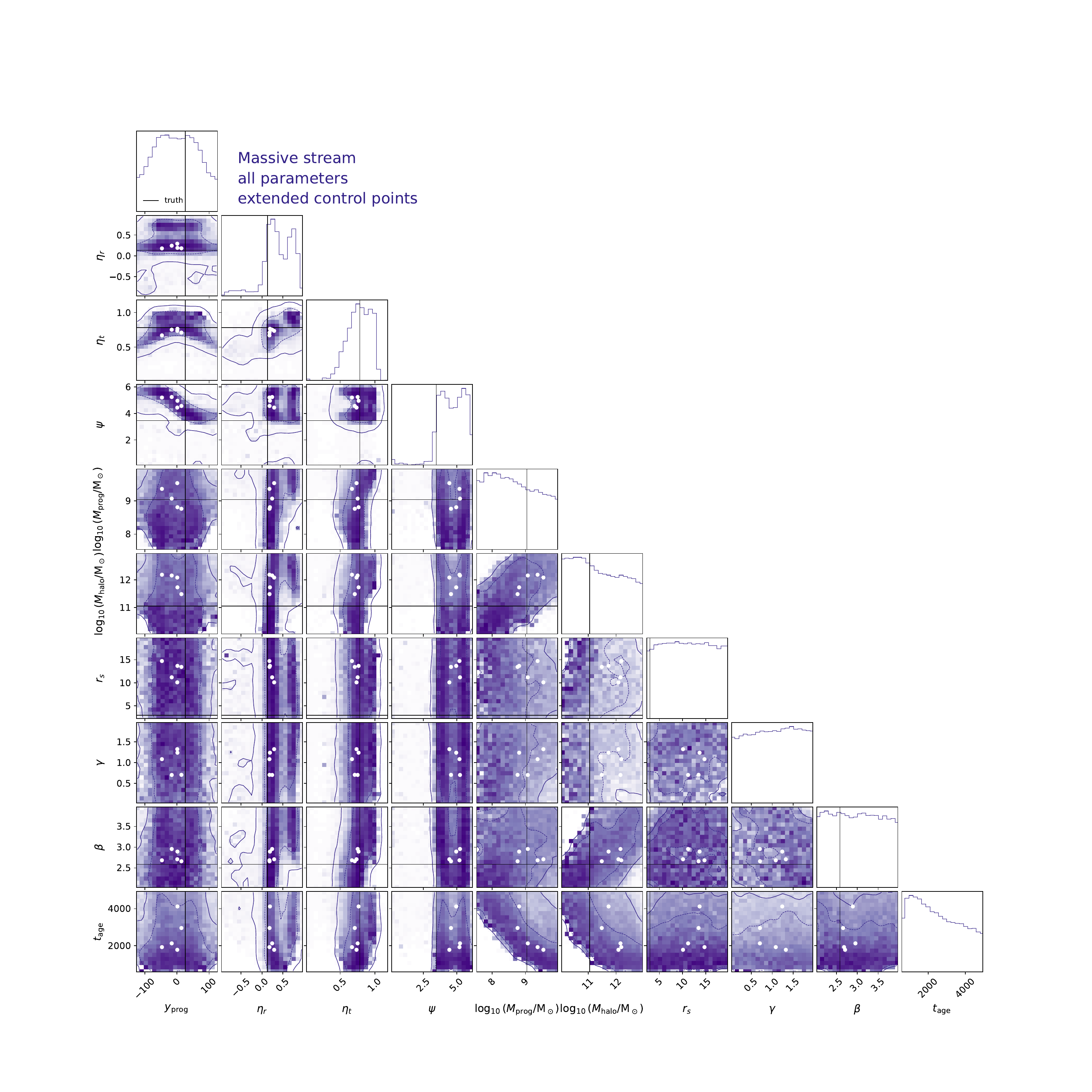}
    \caption{Constraints on the 10 free parameters for the Massive stream, but running \texttt{X-Stream} with the extended red control points shown in Figure \ref{fig:extendedpoints} (left) as input data. 
    The lines and points are the same as in Figure \ref{fig:corner_inner}.
    }
    \label{fig:corner_massive_extendedpoints}
\end{figure*}

Figure \ref{fig:corner_massive_extendedpoints} shows the constraints for the Massive stream \texttt{X-Stream} run. The lines and points are the same as in Figure \ref{fig:corner_inner}.

Including the extended debris reshapes the $y_{\rm prog}$-$\eta_r$ joint posterior. Where Figure \ref{fig:corner_inner} showed a bimodal $y_{\rm prog}$ posterior with $\eta_r$ only bounded from below, here the $y_{\rm prog}$ posterior is flat across most of the prior range, while $\eta_r$ now forms a closed, bimodal joint posterior instead of running into the upper prior boundary. The true $\eta_r$ and $\psi$ values remain at the lower edge of their posteriors, as in the non-extended run, so the extended debris sharpens the shape of the constraint without moving the truth relative to it.

The $M_{\rm halo}$-$M_{\rm prog}$ joint posterior shows a similar strong, open degenerate band as in Figure  \ref{fig:corner_inner}, still containing the true halo and progenitor mass, and still without a closed constraint on either mass. Unlike the S-shaped stream, where extended debris tightens the mass constraint (Section \ref{sec:extendeddebrisdiscus}), the extended control points do not break this degeneracy for the Massive stream.

The remaining parameters show the same general trends as in the non-extended run. $r_s$ and $\gamma$ remain weakly constrained across the full prior range, with no closed joint posterior against either parameter. The joint constraint between 
$\beta$ and $M_{\rm halo}$, shows  the  degenerate trend discussed above, but is now tightened as compared to the run without extended control points. The joint constraints between $t_{\rm age}$ and $M_{\rm halo}$, and  $t_{\rm age}$ and $M_{\rm prog}$ are also tighter in this run with more extended debris: shorter integration times pair with higher progenitor and halo masses. \citet{NibPear2025} discuss the same degeneracies, showing that a lower-mass progenitor needs a longer integration time to produce a stream of the same length, since its stars escape at lower velocities, and that a lower-mass halo requires a longer orbital time to produce a sufficiently long stream, since its debris phase mixes more slowly.

\begin{figure*}
    \centering
    \includegraphics[width=\textwidth]{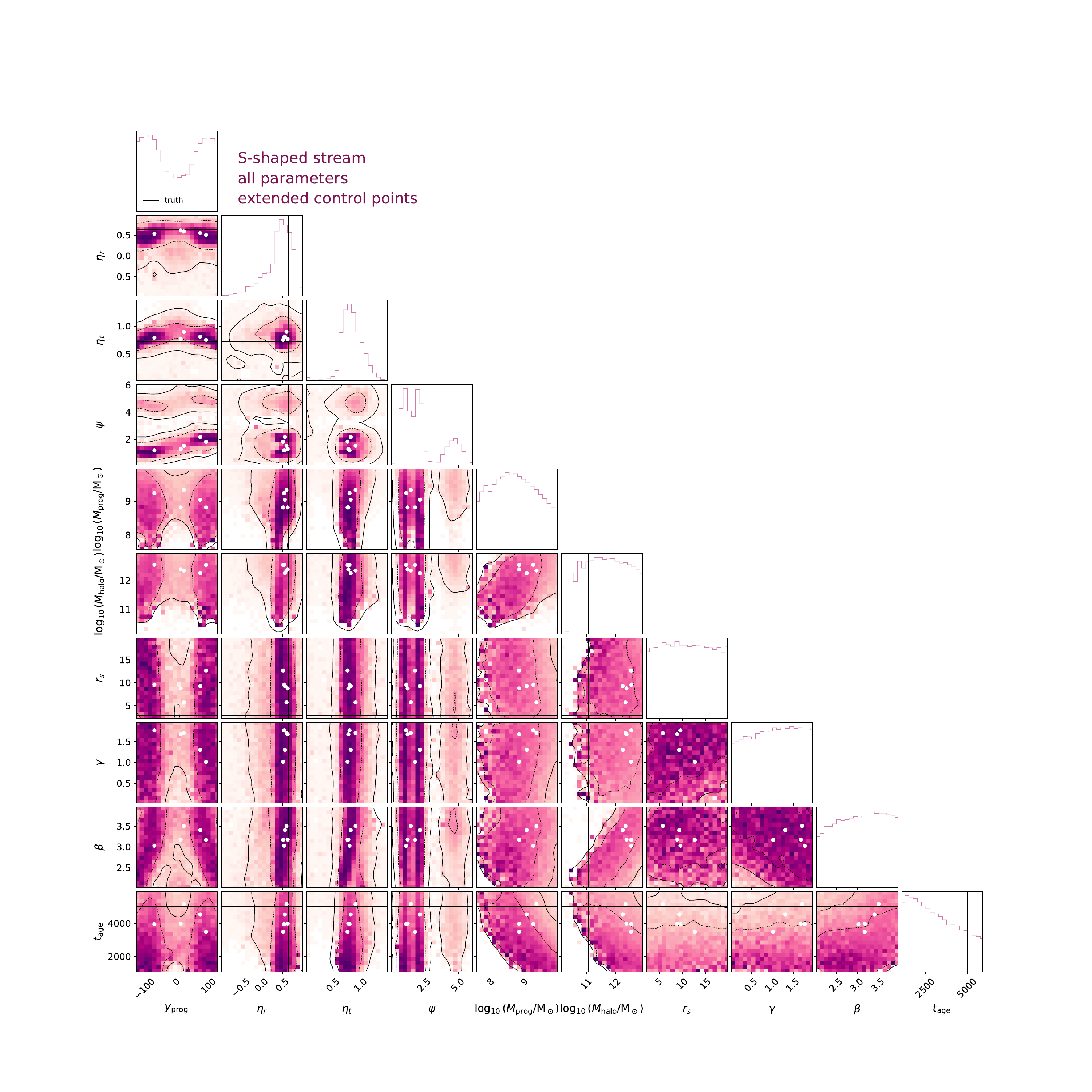}
    \caption{Constraints on the 10 free parameters for the S-shaped stream, but running \texttt{X-Stream} with the extended red control points shown in Figure \ref{fig:extendedpoints} (right) as input data.  The lines and points are the same as in Figure \ref{fig:corner_inner}.   
    }
    \label{fig:corner_sshaped_extendedpoints}
\end{figure*}

Lastly, Figure \ref{fig:corner_sshaped_extendedpoints} shows the constraints for the S-shaped stream \texttt{X-Stream} run using the extended control points from Figure \ref{fig:extendedpoints}. The lines and points are the same as in Figure \ref{fig:corner_inner}.

Including the extended debris closes the orbital joint posteriors. Where Figure \ref{fig:corner_outer} showed $\eta_r$ and $\eta_t$ each bounded only from below and strong constraints on $\psi$, here all three orbital parameters form closed contours, with the true values recovered within the 68\% region. The $y_{\rm prog}$ posterior remains bimodal with the truth contained within the 68\% region in the positive peak. 

The $M_{\rm halo}$-$M_{\rm prog}$ joint posterior also changes shape: the strong degeneracy persists, but the extended debris now sets a lower limit on halo mass, unlike Figure \ref{fig:corner_outer}, where the band was unbounded from below. We still do not see an upper limit on halo mass. This tightening is consistent with there being fewer ways to reproduce the correct morphology once the control points trace the wrap with its specific curvature (Section \ref{sec:extendeddebrisdiscus}).

The remaining parameters follow the same pattern as for the Massive stream. $r_s$ and $\gamma$ remain weakly constrained across the full prior range in both runs. The joint constraint between $\beta$ and $M_{\rm halo}$ preserves the same degenerate trend discussed above, but is tighter with the extended control points. The joint constraints between $t_{\rm age}$ and $M_{\rm halo}$, and between $t_{\rm age}$ and $M_{\rm prog}$, are likewise tighter, with shorter integration times pairing with higher progenitor and halo masses, consistent with the degeneracies discussed in \citet{NibPear2025}.

\bibliography{sample631}{}
\bibliographystyle{aasjournal}

\end{document}